\documentclass[12pt, draftclsnofoot, onecolumn]{IEEEtran}

\newtheorem{theo}{Theorem}

\newtheorem{cor}{Corollary}

\IEEEoverridecommandlockouts

\makeatletter
\newcommand{\biggg}[1]{{\hbox{$\left#1\vbox to 20.5pt{}\right.\n@space$}}}
\newcommand{\Biggg}[1]{{\hbox{$\left#1\vbox to 23.5pt{}\right.\n@space$}}}
\newcommand{\bigggg}[1]{{\hbox{$\left#1\vbox to 26.5pt{}\right.\n@space$}}}
\newcommand{\Bigggg}[1]{{\hbox{$\left#1\vbox to 29.5pt{}\right.\n@space$}}}
\newcommand{\biggggg}[1]{{\hbox{$\left#1\vbox to 32.5pt{}\right.\n@space$}}}
\newcommand{\Biggggg}[1]{{\hbox{$\left#1\vbox to 35.5pt{}\right.\n@space$}}}
\newcommand{\bigggggg}[1]{{\hbox{$\left#1\vbox to 38.5pt{}\right.\n@space$}}}
\newcommand{\Bigggggg}[1]{{\hbox{$\left#1\vbox to 41.5pt{}\right.\n@space$}}}
\makeatother

\usepackage{bm,cite,algorithm,algorithmic,float,amsmath,amssymb}
\usepackage[dvips]{graphicx}
\usepackage{subfigure,amsfonts}
\usepackage{mdwmath}
\usepackage{mdwtab}
\usepackage{xcolor,color}
\usepackage{cool}
\usepackage{balance}
\usepackage{diagbox}
\floatname{algorithm}{Algorithm}

\makeatletter
\renewcommand\paragraph{\@startsection{paragraph}{4}{\z@}%
            {-2.5ex\@plus -1ex \@minus -.25ex}%
            {1.25ex \@plus .25ex}%
            {\normalfont\normalsize\itshape}}
\makeatother
\usepackage{epstopdf}

\usepackage[style=0]{mdframed}
\usepackage{showexpl}

\mdfdefinestyle{exampledefault}{%
rightline=true,innerleftmargin=10,innerrightmargin=10,
frametitlerule=true,frametitlerulecolor=green,
frametitlebackgroundcolor=yellow,
frametitlerulewidth=2pt}

\allowdisplaybreaks

\begin{document}

\markboth{SUBMITTED TO IEEE JOURNAL ON SELECTED AREAS IN COMMUNICATIONS}{Liu \MakeLowercase{\textit{et al.}}: Rate Splitting Multiple Access for Semi-Grant-Free Transmissions}

\title{Rate Splitting Multiple Access for Semi-Grant-Free Transmissions}

\author{
Hongwu~Liu, Theodoros A. Tsiftsis,  Bruno Clerckx,  
Kyeong Jin Kim,\\
Kyung Sup Kwak, and H. Vincent Poor, \IEEEmembership{Life Fellow, IEEE}

\thanks{H. Liu is with the School of Information Science and Electrical Engineering, Shandong Jiaotong University, Jinan 250357, China (e-mail: liuhongwu@sdjtu.edu.cn).}
\thanks{T. A. Tsiftsis is with the School of Intelligent Systems Science and Engineering, Jinan University, Zhuhai 519070, China (e-mail: theo\_tsiftsis@jnu.edu.cn).}
\thanks{K. J. Kim is with Mitsubishi Electric Research Laboratories, Cambridge, MA 02139 USA (e-mail: kkim@merl.com).}
\thanks{B. Clerckx is  with the Department of Electrical and Electronic Engineering, Imperial College London, London, UK (e-mail: b.clerckx@imperial.ac.uk).}
\thanks{K. S. Kwak is with the Department of Information and Communication Engineering, Inha University, Incheon 22212, South Korea (e-mail: kskwak@inha.ac.kr).}
\thanks{H. V. Poor is with the Department of Electrical and Computer Engineering, Princeton University, Princeton, NJ 08544 USA (e-mail: poor@princeton.edu).}
}

\maketitle
\setcounter{page}{1}
\begin{abstract}
Enabled by hybrid grant-based (GB) and grant-free (GF) transmission techniques, 
GF users of internet of things (IoT) devices and massive machine-type communications (mMTC) meet opportunities to share wireless resources with GB users. In this paper, we propose a rate splitting multiple access (RSMA) strategy for an emerging semi-grant-free  (SGF) transmission system to increase connectivity and reliability. In the proposed RSMA assisted SGF (RSMA-SGF) scheme, the GF users apply the rate splitting principle to realize distributed contentions and utilize transmit power most effectively for robust transmissions, meanwhile keeping themselves transparent to the GB user. Compared to existing non-orthogonal multiple access (NOMA) assisted SGF schemes, the RSMA-SGF scheme significantly decreases outage probability and achieves full multiuser diversity gain without restricting the GB and GF users' target rates to a limited value region. Exact expressions and asymptotic analysis for the outage probability are provided to facilitate the system performance evaluation of the proposed RSMA-SGF scheme. Computer simulation results clarify the superior outage performance of the RSMA-SGF scheme and verify the accuracy of the developed analytical results. 
\end{abstract}

\begin{IEEEkeywords}
Rate splitting, multiple access, grant-free transmissions, outage probability.
\end{IEEEkeywords}

\section{Introduction}

The explosive growth of internet-of-things (IoT) devices, intelligent robotics, and Industry 4.0 networks has created unprecedented demands for massive access, heterogeneous mobile traffic, and highly spectral efficient connectivity. On the road map of the fifth-generation (5G) evolution, enhanced mobile broadband (eMBB), ultra-reliable low latency communications (URLLC), and massive machine-type communications (mMTC) were standardized to facilitate the proliferation of the emerging application domains. Grant-free (GF) transmissions, which represent ubiquitous scenarios of beyond 5G URLLC and short packet traffic emitted by various sensors and massive IoT devices, have been envisioned to be a new air-interface trend for the next generation IoT \cite{Next_IoT, Sparse_GF_IoT, SCMA_blind}. Compared to conventional grant-based (GB) transmissions, where the amount of handshaking signalling could exceed the amount of data sent by devices, GF transmissions grant devices access without lengthy handshaking protocols \cite{SCMA_blind}. Taking advantages of spectrum sharing from non-orthogonal multiple access (NOMA), semi-grant-free (SGF) transmissions opportunistically admitted GF users on the resource blocks occupied by the GB users in a multiplexing way, which surprisingly results in a higher spectrum efficiency than purely admitting GF or GB users \cite{SGF_NOMA_SGMA,SGF_NOMA_Simple}. For NOMA assisted SGF (NOMA-SGF) transmissions, a series of the complicated handshaking process was omitted for granting latency-critical GF user, while collisions from GF users' contention can be elaborately relieved by using the NOMA philosophy\footnote{In this paper, we only consider uplink SGF transmission scenarios. Therefore, all the mentioned NOMA and rate splitting multiple access (RSMA) in what follows are referred to the corresponding uplink schemes or scenarios.} \cite{SGF_NOMA_Simple}. 

Taking into account the quality of service (QoS) requirements of GB users, GF users' contention needs to be carefully controlled for the NOMA-SGF schemes to prevent the system performance degradation to GB users. In particular, the system performance experienced by GB users should be the same as their counterparts in orthogonal multiplex access (OMA). It has been shown that the distributed contention protocol can ensure a fixed number of GF users to be granted access, while the open-loop protocol still suffers from user collisions as that in pure GF transmissions \cite{SGF_NOMA_Simple}. Furthermore, the equivalent received power at the base-station (BS) corresponding to GB transmission was broadcasted to GF users to facilitate distributed contentions \cite{SGF_NOMA_QoS}. To maintain low-latency and high-reliability trade-off in addition to making GF users transparent to GB users, advanced code-domain or power-domain multiplexing and successive interference cancellation (SIC) techniques were exploited for the NOMA-SGF schemes \cite{SGF_NOMA_QoS,SGF_NOMA_Geometry_Conferece,SGF_NOMA_Advanced,SGF_NOMA_PA}. 
Existing designs of the NOMA-SGF schemes largely depend on the SIC decoding order at the BS receiver and/or transmit power allocation. Based on the relative power levels, the hybrid SIC was applied to decode the desired signals, e.g., the admitted GF user's signal can be decoded at the first stage or the second stage of the SIC to achieve the allowed maximum rate \cite{Unveiling_SIC_PartI}. Associated with the hybrid SIC decoding order, transmit power allocation can be applied to improve the transmission reliability of the NOMA-SGF schemes \cite{SGF_NOMA_Advanced,SGF_NOMA_PA}. 

\subsection{Related Work}

The primary concept of NOMA-SGF was found in \cite{SGF_NOMA_SGMA} and \cite{SGF_NOMA_Simple}, respectively. In \cite{SGF_NOMA_SGMA}, the authors discussed the feasibility of applying NOMA to realize the SGF transmissions, including several NOMA solutions such as sparse-code multiple access (SCMA), power-domain NOMA,  pattern-division multiple access (PDMA), and interleave division multiple access (IDMA). Two basic NOMA-SGF schemes were firstly proposed using the power-domain NOMA philosophy without causing too much performance degradation to  GB users \cite{SGF_NOMA_Simple}. Considering stochastic geometry in outage performance analysis, an accurate interference threshold was adopted for granting GF users \cite{SGF_NOMA_Geometry_Conferece,SGF_NOMA_Geometry}. In \cite{SGF_NOMA_QoS} and \cite{Unveiling_SIC_PartI}, the authors proposed to use the hybrid SIC decoding order to ensure that GB user can experience the same system performance as in OMA, meanwhile aiming at improving the transmission reliability for the granted GF user. Adaptive power allocation methods were designed for the NOMA-SGF schemes using the fixed SIC decoding order \cite{SGF_NOMA_PA} and hybrid SIC decoding order \cite{SGF_NOMA_Advanced}, respectively. In \cite{SGF_NOMA_Performance_UL}, the cognitive radio (CR) principle was applied to manage interference from GF users to a selected GB user. 
For the tactile IoT network where the NOMA-SGF is applied, the authors of \cite{SGF_NOMA_Tactile} proposed a solution to the joint channel assignment and power allocation problem. The ergodic rate analysis was provided in \cite{SGF_NOMA_Ergodic_Rate} and most recently a multi-agent deep reinforcement learning algorithm was proposed to optimize the transmit power pool for the NOMA-SGF systems \cite{SGF_NOMA_MA_DRL}. 
 
\subsection{Motivations and Contributions}

For multiple access channels (MACs), rate splitting (RS) was proposed to achieve the capacity region \cite{Rate_Splitting_MAC}. 
Considering that uplink NOMA scenarios can be regarded as MACs, rate splitting multiple access (RSMA) strategies  were applied to achieve the capacity region for uplink NOMA transmissions \cite{RS_CP_SC,RS_NOMA_UL_fair}. Motivated by the CR principles, we developed the adaptive power allocation method for the RSMA to guarantee the QoS requirements of the paired uplink users \cite{RS_NOMA_UL_fair}. For the uplink single-input multiple-output (SIMO) NOMA systems, the RS scheme was designed to ensure max-min fairness among the NOMA users \cite{RS_NOMA_maxmin_fairness}. The sum-rate of the uplink MACs was maximized with the aid of the RSMA by adjusting the users' transmit power and the BS's decoding order \cite{RS_SUM_RATE}. Very recently, the RS techniques were applied for multiple access in the uplink aerial networks \cite{RS_Aerial_Networks}. 

In this paper, we adopt the same uplink SGF transmission scenario originally introduced in \cite{SGF_NOMA_Simple} and lately investigated in \cite{SGF_NOMA_QoS}. An RS strategy is proposed for the SGF transmission system to realize the RSMA assisted SGF (RSMA-SGF) scheme. We apply the RS techniques at the GF users not only for the distributed contentions but also for the SGF transmissions with the aim of improving the transmission reliability for the admitted GF user. 

The main contributions of this paper are  summarized as follows: 
\begin{itemize}
\item We propose RSMA-SGF and contrast with NOMA-SGF for uplink SGF network deployments. The GF users apply the RS strategy to calculate their achievable rates and determine the corresponding back-off time for the distributed contentions. By using the RS strategy, the admitted GF user's transmission is transparent to the GB user, while the transmit power of the admitted GF user is effectively exploited to enhance its transmission reliability. 
\item We derive exact expression for the outage probability and its high signal-to-noise ratio (SNR) approximation to facilitate the evaluation of the outage performance for the proposed RSMA-SGF scheme. The analytical results reveal that the RSMA-SGF scheme can achieve the full multiuser diversity gain without restricting the GB and GF users' target rates to a limited value region, which is more robust than the existing NOMA-SGF schemes.  
\item We evaluate the accuracy of the developed analytical results and high SNR approximations via computer simulations. The impact of several system parameters on the outage performance is revealed. Compared to existing NOMA-SGF schemes, the proposed RSMA-SGF scheme achieves the best outage performance over the existing NOMA-SGF schemes. In addition, due to the capability of RSMA to achieve the capacity region of MACs, the proposed RSMA-SGF scheme works well for a wider region of the target rates than the existing NOMA-SGF schemes. 
\end{itemize}

The remainder of this paper is organized as follows: Section II and Section III present the system model and RSMA-SGF scheme, respectively; In Section IV, the outage performance of the proposed RSMA-SGF scheme is analyzed and the high SNR approximation for the outage probability is presented; In Section V, simulation results are presented for corroborating the superior outage performance of the RSMA-SGF scheme and Section VI summarizes this work.

\section{System Model}

\begin{figure}[tb]
    \hspace{-0.1in}    
    \subfigure[The system model diagram]{
    \includegraphics[width=3in]{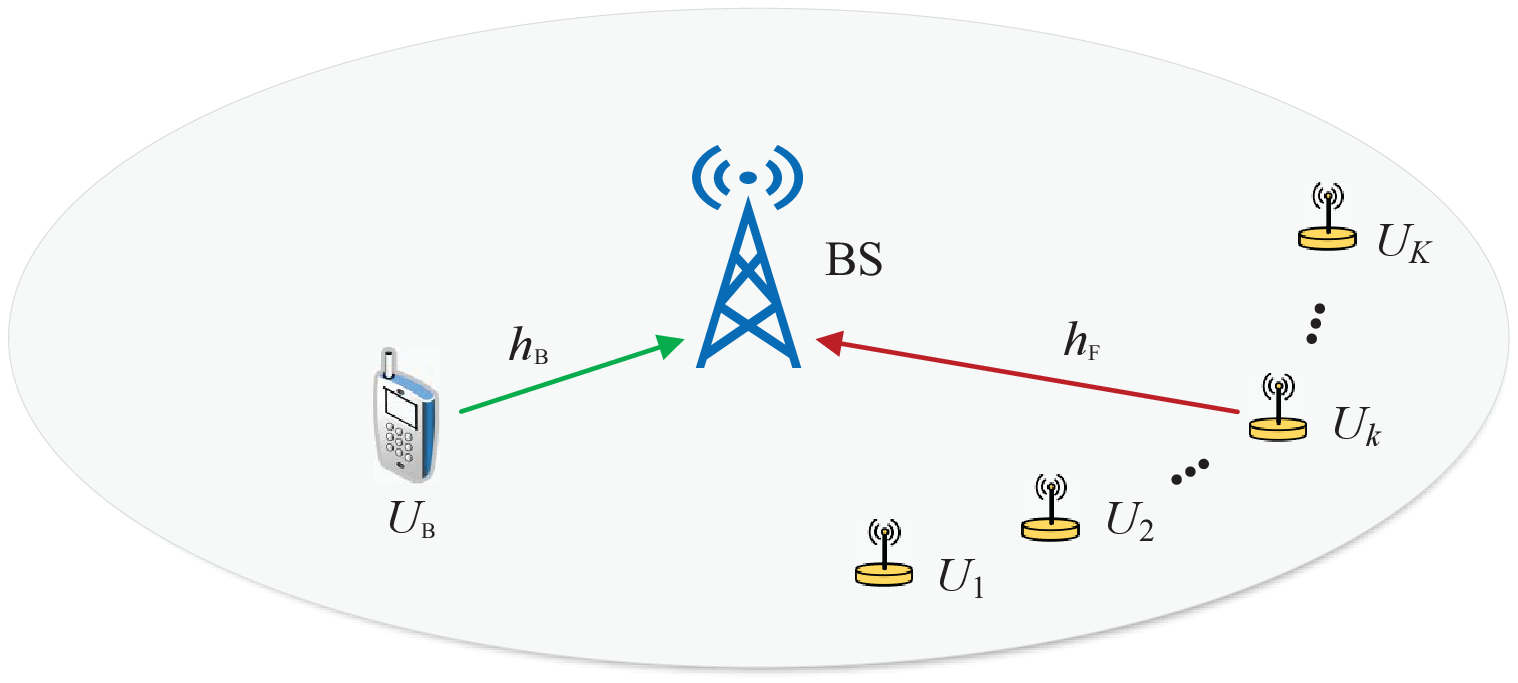}}
    \label{fig:subfig0a}  
    \subfigure[Illustration of the arrived signals at the BS]{
    \includegraphics[width=3.2in]{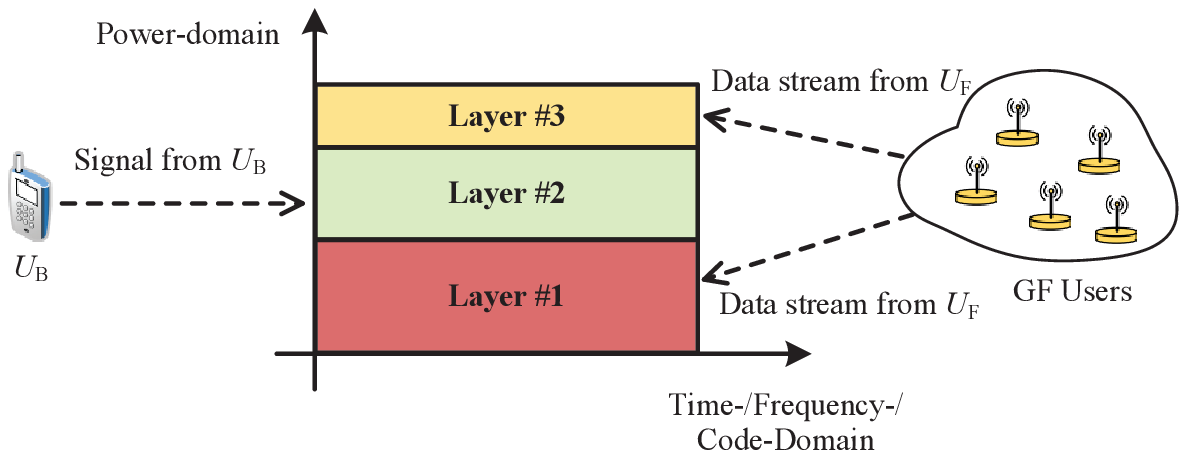}}
    \caption{An illustration of the considered NOMA-SGF system and signal model.}
    \label{fig:subfig0b}    
    \vspace{-0.15in}
\end{figure}

We consider the uplink  scenario in an SGF transmission system, as depicted in Fig. 1(a), where $K$ GF users, $\{U_1, U_2, \cdots, U_K \}$, compete to be paired with a GB user, $U_{_{\rm B}}$, for uplink transmissions. To ensure that the admission of the GF users do not cause too much interference to the GB user, we consider that only a single user is grant access simultaneously with the GB user \cite{SGF_NOMA_QoS}.
After the user contention and access admission, the paired GB and GF users share the same resource block for their uplink transmissions. Denote the GB user's channel by $h_{_{\rm B}}$ and the $k$th $\left(k \in \{1, 2, \ldots, K\}\right)$ GF user's channel by $h_k$, which are modeled as independent and identically distributed (i.i.d.) Rayleigh fading. Without loss of generality, we assume that the GF users' channel gains are ordered as
\begin{eqnarray}
|h_1|^2 \le |h_2|^2 \le \cdots \le |h_K|^2     \label{eq:h_order}
\end{eqnarray}
to facilitate the system performance analysis. 
It should be noted that the above ordering information is unavailable to the BS and all the GF users. In addition, we assume that the GB and GF users have the knowledge of their own CSI, respectively, and the BS has acquired the knowledge of the CSI of the GB user as well as its transmit power, $P_{_{\rm B}}$. In addition, the CSI of the admitted GF user is not required to be known at the BS prior to transmission.  The detailed CSI acquisition procedure will be discussed in Subsection B of Section III. We assume that the channels follow a statistical fading, i.e., the channel fading coefficients keep constant during a single transmission block and can vary from one transmission block to another independently. 

\section{Rate Splitting Multiple Access Assisted Semi-Grant-Free Scheme}

In the proposed RSMA-SGF scheme, the grant access of the GF user is transparent to the GB user. Therefore, we only consider that the RS is carried out at the GF user and the RSMA-SGF scheme is designed to improve the achievable rate of the GF user over its counterpart in a NOMA-SGF scenario where the RS is not applied.

\subsection{Signal Model}

We assume that the GF user $U_{_{\rm F}} \in \{U_1, U_2, \cdots, U_K \}$ is admitted to transmit its signal $x_{\rm F}$ being paired with the GB user. 
Before the uplink transmission, $U_{_{\rm F}}$ splits its signal $x_{_{{\rm F}}}$ into two streams $x_{_{{\rm F},1}}$ and $x_{_{{\rm F},2}}$. The transmit power of the admitted GF user,  $P_{_{\rm F}}$, is allocated for the two streams according to $\tilde x_{_{{\rm F}}} = \sqrt{\alpha P_{_{\rm F}}}x_{_{{\rm F},1}} + \sqrt{(1-\alpha)P_{_{\rm F}}}x_{_{{\rm F},2}}$, where $\alpha$ is the transmit power allocation factor satisfying $0 \le \alpha \le 1$. In each transmission block, the paired GB and GF users transmit their signals simultaneously to the BS and the received signal at the BS can be written as
\begin{eqnarray}
y = \sqrt{P_{_{\rm B}}} h_{_{\rm B}} x_{_{\rm B}} + \sqrt{\alpha P_{_{\rm F}}} h_{_{\rm F}} x_{_{{\rm F},1}} + \sqrt{(1-\alpha)P_{_{\rm F}}} h_{_{\rm F}} x_{_{{\rm F},2}} + w, \label{eq:y}
\end{eqnarray}
where $x_{_{\rm B}}$ is the GB user's signal, $h_{_{\rm F}} \in \{ h_1, h_2, \cdots,$ $h_K \}$ is the channel of the admitted GF user, and $w$ is additive white Gaussian noise (AWGN) with zero mean and unit variance. We assume that the transmitted signal $x_i \in \{x_{_{\rm B}}, x_{_{\rm F}}, x_{_{{\rm F},1}}, x_{_{{\rm F},2}} \}$ is obtained by independent coding with Gaussian code book satisfying ${\mathbb{E}}\{|x_i|^2\} = 1$. 
As an example, we give an illustration in Fig. 1(b) for the arrived power levels at the BS, where Layers $\#1$, $\#2$, and $\#3$ correspond to $\sqrt{\alpha P_{_{\rm F}}} h_{_{\rm F}} x_{_{{\rm F},1}}$, $\sqrt{P_{_{\rm B}}} h_{_{\rm B}} x_{_{\rm B}} $, and $\sqrt{(1-\alpha)P_{_{\rm F}}} h_{_{\rm F}} x_{_{{\rm F},2}}$, respectively. In a nutshell, Layers $\#1$, $\#2$, and $\#3$ are superimposed in the power-domain according to the RSMA philosophy \cite{Rate_Splitting_MAC}.  

In the signal model of \eqref{eq:y}, the RS is carried out only for $0< \alpha <1$. For the case of $\alpha = 0$ or $\alpha =1$, the admitted GF user transmits the signal $ x_{_{{\rm F},2}} = x_{_{{\rm F}}}$ or $ x_{_{{\rm F},1}} = x_{_{{\rm F}}}$ without operating in the RS mode. Consequently, the SIC decoding modes at the BS are determined by the different values of $\alpha$,  which are discussed as follows:

\begin{itemize}
\item For the case of $0<\alpha<1$, the decoding order $x_{_{{\rm F},1}} \to x_{_{\rm B}} \to x_{_{{\rm F},2}}$ is applied at the BS to recover the desired signals, $x_{_{{\rm F},1}}$, $x_{_{\rm B}}$, and $x_{_{{\rm F},2}}$, sequentially \cite{RS_NOMA_UL_fair}. In other words, the GB user's signal is decoded in the second stage of the SIC processing, whereas the admitted GF user's streams are decoded in the first and third stages, respectively. Once the streams $x_{_{{\rm F},1}}$ and $x_{_{{\rm F},2}}$ are recovered successfully, all the information transmitted by the GF user can be retrieved at the BS. In this case, the achievable rates for transmitting  $x_{_{{\rm F},1}}$, $x_{_{\rm B}}$, and $x_{_{{\rm F},2}}$ can be expressed as
\begin{eqnarray}
    R_{{_{{\rm F},1}}} = \log_2\left( 1+ \frac{\alpha P_{_{\rm F}} |h_{_{\rm F}}|^2}{ P_{_{\rm B}}|h_{_{\rm B}}|^2 + (1-\alpha)P_{_{\rm F}} |h_{_{\rm F}}|^2 + 1} \right),   \label{eq:rate_rs1}
\end{eqnarray}
\begin{eqnarray}
    R_{{_{\rm B}}} = \log_2 \left( 1 + \frac{P_{_{\rm B}}|h_{_{\rm B}}|^2}{(1-\alpha)P_{_{\rm F}} |h_{_{\rm F}}|^2 + 1}\right),
\end{eqnarray}
and 
\begin{eqnarray}
    R_{{_{{\rm F},2}}} =  \log_2\left(1 + (1-\alpha)P_{_{\rm F}} |h_{_{\rm F}}|^2  \right),
\end{eqnarray}
respectively. For the admitted GF user, its achievable rate is given by $R_{_{\rm F}} = R_{{_{{\rm F},2}}} + R_{{_{{\rm F},2}}}$. 
\item  For the case of $\alpha=0$, since the admitted GF user only transmits $ x_{_{{\rm F},2}} = x_{_{{\rm F}}}$, the decoding order $x_{_{\rm B}} \to x_{_{{\rm F}}}$ is applied at the BS to recover the desired signals, $x_{_{\rm B}}$ and $x_{_{{\rm F}}}$, sequentially. 
The achievable rates of the paired GB and GF users are given by
$
    R_{{_{\rm B}}} = \log_2 \left( 1 + \frac{P_{_{\rm B}}|h_{_{\rm B}}|^2}{ P_{_{\rm F}} |h_{_{\rm F}}|^2 + 1}\right)
$
and 
$
    R_{{_{{\rm F}}}} =  \log_2\left(1 +  P_{_{\rm F}} |h_{_{\rm F}}|^2  \right)
$, respectively. 
\item For the case of $\alpha=1$, the admitted GF user only transmits $ x_{_{{\rm F},1}} = x_{_{{\rm F}}}$, the decoding order $x_{_{\rm F}} \to x_{_{{\rm B}}}$ is applied at the BS to recover the desired signals, $x_{_{\rm F}}$ and $x_{_{{\rm B}}}$, sequentially. 
The achievable rates of the paired GF and GB users are given by
$
    R_{{_{\rm F}}} = \log_2 \left( 1 + \frac{P_{_{\rm F}}|h_{_{\rm F}}|^2}{ P_{_{\rm B}} |h_{_{\rm B}}|^2 + 1}\right)
$
and 
$
    R_{{_{{\rm B}}}} =  \log_2\left(1 +  P_{_{\rm B}} |h_{_{\rm B}}|^2  \right)
$,
respectively. 
\end{itemize}

\subsection{RSMA-SGF Scheme}

Prior to admission, the BS broadcasts an interference threshold $\tau$ to all the $K$ GF users to realize the distributed contention.
Recalling that the GB user's signal needs to be decoded correctly in the second stage of the SIC processing for $0 < \alpha < 1$, it requires that $\log_2 \left( 1 + \frac{P_{_{\rm B}}|h_{_{\rm B}}|^2}{(1-\alpha)P_{_{\rm F}} |h_{_{\rm F}}|^2 + 1}\right) \ge \hat R_{_{\rm B}}$ with $\hat R_{_{\rm B}}$ denoting the GB user's target transmission rate. Therefore, the interference threshold is designed as $\tau = \max\left\{0, \hat\tau\left(|h_{_{\rm B}}|^2\right)\right\}$, where $\hat \tau\left(|h_{_{\rm B}}|^2\right) = \tfrac{P_{_{\rm B}}|h_{_{\rm B}}|^2}{2^{\hat R_{_{\rm B}}}-1} - 1$  denotes the maximum interference power allowed by the GB user's QoS requirements. 
In the proposed RSMA-SGF scheme, the GF user that achieves the allowed maximum $R_{_{\rm F}}$  will be admitted for transmissions. 

The admission procedure of the RSMA-SGF scheme consists of the following steps:
\begin{itemize}
    \item The BS broadcasts pilot signal to assist the GB and GF users to estimate their channels.    
    \item The GB user feeds back its transmission power $P_{_{\rm B}}$ and CSI $h_{_{\rm B}}$ to the BS. 
    \item The BS calculates the interference threshold $\tau$ and broadcasts it to all the $K$ GF users. 
	\item Each GF user calculates its achievable rate and determines the associated backoff time. 
	\item Through the distributed contention, the GF user that has the minimum backoff time is admitted.   
\end{itemize}

According to the relationship between $P_{_{\rm F}} |h_k|^2$ and $\tau$,  the $K$ GF users are categorized into two groups and determine their achievable rates as in the following:
\begin{itemize}
    \item Group I: In this group, the received signal power at the BS corresponding to the $U_k$'s transmission is less than or equal to $\tau$, i.e., $P_{_{\rm F}} |h_k|^2 \le \tau$. To achieve the allowed maximum transmission rate, $U_{k}$ does not conduct RS and transmits its signal directly by setting $\alpha = 0$. At the BS receiver, the decoding for the GB user's signal is carried out at the first stage of the SIC processing, while the decoding for the $U_k$'s signal is carried out at the second stage, which yields the achievable rate $R_{_{{\rm I},k}} =  \log_2\left(1 + P_{_{\rm F}} |h_{k}|^2  \right)$ for the $U_k$'s transmission.  
    \item Group II: In this group, the received signal power at the BS corresponding to the $U_k$'s transmission is larger than $\tau$, i.e., $P_{_{\rm F}} |h_k|^2 > \tau$. For the GF user $U_k$ in this group, RS is carried out for its transmission by setting $\alpha = 1 - \tfrac{\tau}{P_{_{\rm F}}|h_k|^2}$, which yields the achievable rate $\log_2(1 + \tau)$ for transmitting $x_{_{{\rm F},2}}$. Substituting $\alpha = 1 - \tfrac{\tau}{P_{_{\rm F}}|h_k|^2}$ into \eqref{eq:rate_rs1}, the achievable  rate for transmitting $x_{_{{\rm F},1}}$ can be determined and the total achievable rate for the $U_k$'s transmission can be written as $R_{_{{\rm II},k}} =  \log_2\left( 1 + \frac{P_{_{\rm F}}|h_k|^2 - \tau}{P_{_{\rm B}}|h_{_{\rm B}}|^2 + \tau + 1} \right) +  \log_2(1 + \tau)$. 
    \end{itemize}

Based on the above discussions, the achievable rate of $U_k$ is given by
\begin{eqnarray}
    R_{k}  = \left\{ {\begin{array}{*{20}{c}}
        R_{_{{\rm I},k}},&P_{_{\rm F}} |h_k|^2 \le \tau\\
        R_{_{{\rm II},k}},&P_{_{\rm F}} |h_k|^2 > \tau
        \end{array}} \right., 
\end{eqnarray}
In order to guarantee that the GB user can attain the same outage performance as in OMA, the admitted GF user in Group II is allowed to transmit its signal only when $R_k \ge \hat R_{_{\rm F}} $, where $\hat R_{_{\rm F}}$ denotes the target transmissions rate of all the GF users; Otherwise, the GF user in Group II keeps silent during the transmissions. In this way, the failure of decoding $x_{_{{\rm F},1}}$ in the first stage of the SIC processing can be avoided, while the GB user still experiences the same performance as in OMA. To realize distributed contention, the backoff time of $U_k$ is set to be inversely proportional to its achievable rate \cite{Carrier_Sensing,Cooperative_Path_Selection,Feedback_Reliable_Selection}. Thus, the GF user with the largest $R_k$ will be granted access in a distributed manner. 

{\textbf{\emph{Remark 1}}}: Since $ R_{_{{\rm I},k}}$ is  monotonically increasing with respect to $|h_k|^2$ while $|h_1|^2 \le |h_2|^2 \le \cdots \le |h_K|^2$, the GF user having the largest $|h_k|^2$ in Group I will be granted access if Group II is empty. Similarly, the GF user having the largest $|h_k|^2$ in Group II will be granted access if Group I is empty. When both Group I and Group II are not empty, a single user has the chance of being granted access since that $R_{_{{\rm II},k}} > R_{_{{\rm I},k}}$ always holds based on the fact $R_{_{{\rm I},k}} \le \log_2(1 + \tau)$. 
Therefore, in the proposed RSMA-SGF scheme, only a single user who has the largest $|h_k|^2$ will be granted access and the backoff time of all the $K$ GF users can be directly set to be inversely proportional to their channel gains, respectively. 

{\textbf{\emph{Remark 2}}}: If all the GF users are in Group I under the condition that  $\tau > 0$, the admitted GF user will set $\alpha = 0$. When $\tau = 0$ occurs, i.e., the GB user cannot guarantee its QoS requirements, all the GF users are in Group II and the admitted GF user will set $\alpha = 1$. For the cases of $\alpha=0$ and $\alpha = 1$, the admitted GF user does not carry out the RS for the transmissions and transmits its signal $x_{_{{\rm F}}} $ directly. Besides, $R_{_{{\rm II},k}} =  \log_2\left( 1 + \frac{P_{_{\rm F}}|h_k|^2 }{P_{_{\rm B}}|h_{_{\rm B}}|^2  + 1} \right)$ when $\tau = 0$ occurs. 

{\textbf{\emph{Remark 3}}}: If a GF user in Group II is to be admitted, it keeps silent during the transmissions when $R_{_{\rm F}} < \hat R_{_{\rm F}}$. In this case, although the GF user suffers an outage event due to its silent behavior, the GB user occupies the granted resource blocks solely, i.e., the GB user transmits its signal as in an OMA scenario. Recalling that the admitted GF user in Group I also does not affect the outage performance of the GB user compared to that in OMA, the proposed RSMA-SGF scheme results in the same outage performance for the GB user as in OMA.

\section{Outage Performance}

In the proposed RSMA-SGF scheme, the admission of the GF user is transparent to the GB user whose performance is the same as in OMA. Thus, we mainly focus on the outage performance of the admitted GF user.  In this section, we derive closed-form analytical expression  for the outage probability of the admitted GF user. Then, its asymptotic outage performance is investigated in the high SNR region.  

To characterize the outage performance, we denote the event that there are $k$ GF users in Group I by $E_k$, which can be explicitly expressed by 
\begin{eqnarray}
    E_k = \left\{ |h_k|^2 \le \frac{\tau}{P_{_{\rm F}}}, |h_{k+1}|^2 > \frac{\tau}{P_{_{\rm F}}} \right\}
\end{eqnarray} 
for $1 \le k \le K-1$. Furthermore, we denote the two extreme events with no user in Group I and Group II by $E_0 = \left\{ |h_1|^2 > \tfrac{\tau}{P_{_{\rm F}}} \right\}$ and $E_K = \left\{ |h_K|^2 < \tfrac{\tau}{P_{_{\rm F}}} \right\}$, respectively.

When the event $E_k$ ($1 \le k \le K-1$) occurs, Group II contains the $K-k$ GF users among them who has the largest achievable rate will be admitted. For this case, the achievable rate of the admitted GF user is given by $\max\{ R_{_{{\rm II},m}}, k  <  m  \le  K\} $. If $\max\{ R_{_{{\rm II},m}}, k  <  m  \le  K\} < \hat R_{_{\rm F}} $, an outage event happens for $E_k$. As such, the outage events for $E_0$ and $E_K$ can be defined. Therefore, the outage probability experienced by the admitted GF user can be expressed as
\begin{eqnarray}
    P_{\rm out}  &\!\!\!=\!\!\!&  \Pr \left( E_0, \max\{ R_{_{{\rm II},m}}, 1\! \le \!m \!\le\! K\} \!<\! \hat R_{_{\rm F}} \right) +  \sum\limits_{k=1}^{K-1} \Pr \left( E_k, \max\{ R_{_{{\rm II},m}}, k \!<\! m \!\le\! K\} \!<\! \hat R_{_{\rm F}} \right)    \nonumber \\
    &\!\!\!\!\!\! & + \Pr \left( E_K, \max\{ R_{_{{\rm I},m}}, 1 \!\le\! m \!\le\! K\} < \hat R_{_{\rm F}} \right).
\end{eqnarray}
Because the grant-free users’ channel gains are ordered as in \eqref{eq:h_order} while $ R_{_{{\rm I},k}}$ and  $ R_{_{{\rm II},k}}$ are monotonically increasing with respect to $|h_k|^2$, the outage probability can be simplified as follows:
\begin{eqnarray}
    P_{\rm out}  &\!\!\!=\!\!\!&  \Pr \left( E_0,  R_{_{{\rm II},K}} < \hat R_{_{\rm F}} \right)  +  \sum\limits_{k=1}^{K-1} \Pr \left( E_k,  R_{_{{\rm II},K}}  < \hat R_{_{\rm F}} \right)    +  \Pr \left( E_K,  R_{_{{\rm I},K}} < \hat R_{_{\rm F}} \right) .
    \label{eq:Pout_simplified0}
\end{eqnarray}
Define $\epsilon_{_{\rm B}} \triangleq 2^{\hat R_{_{\rm B}}}-1$, $\epsilon_{_{\rm F}} \triangleq 2^{\hat R_{_{\rm F}}} - 1  $, $\eta_{_{\rm B}} \triangleq \frac{\epsilon_{_{\rm B}}}{P_{_{\rm B}}}$, and  $\eta_{_{\rm F}} \triangleq \frac{\epsilon_{_{\rm F}}}{P_{_{\rm F}}}$, we have  
\begin{eqnarray}
    \tau = \max \left\{ 0,  \frac{|h_{_{\rm B}}|^2}{\eta_{_{\rm B}}} - 1 \right\} = 0    \label{eq:tau_less_0}
\end{eqnarray}
when $|h_{_{\rm B}}|^2 < \eta_{_{\rm B}}$. 
Therefore, the outage probability  experienced by the admitted GF user can be rewritten as follows:
\begin{eqnarray} 
        P_{\rm out}  &\!\!\!=\!\!\!& \underbrace {\Pr \left( E_0, |h_{_{\rm B}}|^2 > \eta_{_{\rm B}}, R_{_{{\rm II},K}} < \hat R_{_{\rm F}} \right)}_{Q_0}  +  \sum\limits_{k=1}^{K-1} \underbrace {\Pr \left( E_k, |h_{_{\rm B}}|^2 > \eta_{_{\rm B}}, R_{_{{\rm II},K}}  < \hat R_{_{\rm F}} \right)}_{Q_k}  \nonumber \\
        &\!\!\! \!\!\!&  +  \underbrace {\Pr \left( E_K, |h_{_{\rm B}}|^2 > \eta_{_{\rm B}},  R_{_{{\rm I},K}} < \hat R_{_{\rm F}} \right)}_{Q_K} + \underbrace {\Pr \left( |h_{_{\rm B}}|^2 < \eta_{_{\rm B}}, R_{_{{\rm II},K}} < \hat R_{_{\rm F}} \right)}_{Q_{K+1}}.    \label{eq:Pout_GF}
\end{eqnarray}   

For the proposed RSMA-SGF scheme, the following theorem provides an exact expression for the outage probability experienced by the admitted GF user. 

\begin{theo}
    Assume that $K \ge 2$, the outage probability experienced by the admitted GF user is given by
    \begin{eqnarray}
        P_{\rm out}  &\!\!\!=\!\!\!& \frac{\varphi_0}{K(K-1)}  \sum\limits_{\ell=0}^{K}  \binom{K}{\ell}   (-1)^\ell  ~ \mu_{1} \nu(0, \mu_{2})  \nonumber \\
        &\!\!\! \!\!\!& +  \sum\limits_{k=1}^{K-2}  \varphi_k  \sum\limits_{n=0}^{K-k}   \binom{K\!-\!k}{n} (-1)^n \sum\limits_{\ell=0}^{k} \binom{k}{\ell} (-1)^{\ell}e^{\frac{\ell}{P_{_{\rm F}}}}   \mu_3 \nu(\ell,  \mu_4) \nonumber \\
        &\!\!\! \!\!\!& + \frac{ \varphi_0}{K-1} \sum\limits_{\ell=0}^{K-1} \binom{K\!-\!1}{\ell}  (-1)^\ell e^{\frac{\ell}{P_{_{\rm F}}} }     \left( e^{\frac{1}{P_{_{\rm F}}}} \nu(\ell, \mu_5)  - e^{-\frac{\epsilon_{_{\rm B}} + \epsilon_{_{\rm F}} + \epsilon_{_{\rm B}}\epsilon_{_{\rm F}}  }{P_{_{\rm F}}} } \nu(\ell, \mu_6)   \right)  \nonumber \\
        &\!\!\! \!\!\!& + \sum\limits_{\ell = 0}^K \binom{K}{\ell} (-1)^\ell e^{\frac{\ell}{P_{_{\rm F}}}} \nu(\ell, 0)   + \left(1 - e^{-\eta_{_{\rm F}}}  \right)^K e^{- \eta_{_{\rm B}} (1+\epsilon_{_{\rm F}})  }    \nonumber \\
        &\!\!\! \!\!\!& + \sum\limits_{\ell = 0}^K \binom{K}{\ell}  (-1)^\ell e^{-\ell \eta_{_{\rm F}}} \frac{1 - e^{-(1+\ell \eta_{_{\rm F}} P_{_{\rm B}} )\eta_{_{\rm B}} }}{1 + \ell \eta_{_{\rm F}}  P_{_{\rm B}}},
    \end{eqnarray}
    where $\mu_{1} = e^{\frac{K - \ell(1+\epsilon_{_{\rm B}})(1+\epsilon_{_{\rm F}})}{P_{_{\rm F}}}}$, $\mu_{2} = \frac{K-\ell}{P_{_{\rm F}} \eta_{_{\rm B}}} - \frac{P_{_{\rm B}}\ell}{P_{_{\rm F}}}$, $ \mu_3 = e^{\frac{K-k - n \left(1+\epsilon_{_{\rm B}} \right) \left(1+ \epsilon_{_{\rm F}} \right) }{P_{_{\rm F}}}}$, $ \mu_4 = \frac{K-k-n}{P_{_{\rm F}} \eta_{_{\rm B}} }  - \frac{n P_{_{\rm B}}}{P_{_{\rm F}}}  $, $\mu_5 = \frac{1}{P_{_{\rm F}}\eta_{_{\rm B}}}$, $\mu_6 = -\frac{P_{_{\rm B}}}{P_{_{\rm F}}}$, $\varphi_0 = \frac{K!}{(K-2)!}$, $\varphi_k = \frac{K!}{k! (K-k)!}$ for $1 \le k \le K-2$, and 
    \begin{eqnarray}
    \nu(\ell, \mu) = \left\{ {\begin{array}{*{20}{c}}
        {\epsilon_{_{\rm F}}  \eta_{_{\rm B}}  }, &{ {\rm if~~} \mu = -1 - \frac{\ell}{P_{_{\rm F}}\eta_{_{\rm B}}} },\\
        {\frac{e^{- \eta_{_{\rm B}} \left(\frac{\ell  }{P_{_{\rm F}}\eta_{_{\rm B}}}  + \mu + 1\right) } - e^{- \eta_{_{\rm B}} (1+\epsilon_{_{\rm F}})  \left(\frac{\ell  }{P_{_{\rm F}}\eta_{_{\rm B}}}  + \mu + 1\right)  }}{  \frac{\ell }{P_{_{\rm F}}\eta_{_{\rm B}}}  + \mu + 1  } }, &{\rm otherwise}.
        \end{array}} \right.
    \end{eqnarray} 
\end{theo}
\begin{IEEEproof}
    See Appendix A.
\end{IEEEproof}

{\textbf{\emph{Remark 4:}}} Recalling the relationships between the upper and lower bounds on the channel gains for deriving $Q_k$ ($0 \le k \le K$) as in Appendix A, the corresponding upper bound is always larger than the corresponding lower bound when $|h_{_{\rm B}}|^2 <  \eta_{_{\rm B}} (1+\epsilon_{_{\rm F}}) $ without requiring any additional constraints on $\epsilon_{_{\rm B}}$ and $\epsilon_{_{\rm F}}$. Therefore,  the derived expression in Theorem 1 for the outage probability holds for all the feasible values of $\epsilon_{_{\rm B}}$ and $\epsilon_{_{\rm F}}$. 
On the contrary, the outage probability expression derived for the NOMA-SGF scheme as in \cite{SGF_NOMA_QoS} holds only for $\epsilon_{_{\rm B}} \epsilon_{_{\rm F}} < 1 $, which restricts the target rate pair $(\hat R_{_{\rm B}}, \hat R_{_{\rm F}})$ under which the outage floor can be avoided. In the case of $\epsilon_{_{\rm B}} \epsilon_{_{\rm F}} \ge 1 $, the admitted GF user in the NOMA-SGF scheme exhibits an outage floor, whereas this is not the case in the proposed RSMA-SGF scheme. As it will be verified by the simulation results in Section V, the RSMA-SGF scheme outperforms the NOMA-SGF scheme in terms of the outage probability for all the feasible values of $\epsilon_{_{\rm B}}$ and $\epsilon_{_{\rm F}}$. 

{\textbf{\emph{Remark 5:}}} Due to the fact that different $Q_k$s involve different order statistics, the resulted outage probability expression shown in Theorem 1 is complicated. For example, $Q_0$ is a function of three channel gains, $|h_{_{\rm B}}|^2$, $|h_1|^2$, and $|h_K|^2$, whereas $Q_k$ ($1 \le k \le K-2$) is a function of four channel gains, $|h_{_{\rm B}}|^2$, $|h_k|^2$, $|h_{k+1}|^2$, and $|h_K|^2$. Although $|h_{_{\rm B}}|^2$ is independent of $|h_k|^2$ ($1 \le k \le K$), $|h_k|^2$s are dependent order statistics with $1 \le k \le K$, which makes the expression even more involved. Fortunately, insightful approximations at high SNR can be obtained as shown in Theorem 2.

When there is only a single GF user in the considered system, it is paired with the GB user without the distributed contention. In this case, the outage probability experienced by the single GF user can be written as
\begin{eqnarray} 
    P_{\rm out}  &\!\!\!=\!\!\!& \Pr \left( |h_{_{\rm B}}|^2 > \eta_{_{\rm B}},  |h_1|^2 > \tfrac{\tau}{P_{_{\rm F}}},  R_{_{{\rm II},1}} < \hat R_{_{\rm F}} \right)  
    + \Pr \left( |h_{_{\rm B}}|^2 > \eta_{_{\rm B}},  |h_1|^2 < \tfrac{\tau}{P_{_{\rm F}}},  R_{_{{\rm I},1}} < \hat R_{_{\rm F}} \right)  \nonumber \\
    &\!\!\!  \!\!\!&  + \Pr \left( |h_{_{\rm B}}|^2 < \eta_{_{\rm B}},    R_{_{{\rm II},1}} < \hat R_{_{\rm F}} \right).
         \label{eq:Pout_single}
\end{eqnarray}   

By applying the similar steps as in the proof for Theorem 1, the outage probability experienced by a single GF user can be obtained straightforwardly as shown in the following corollary. 

\begin{cor}
    Assume that $K = 1$, the outage probability experienced by the single GF user can be expressed as follows:
    \begin{eqnarray}
        P_{\rm out}  &\!\!\!=\!\!\!& 1 - e^{-\frac{\epsilon_{_{\rm B}} + \epsilon_{_{\rm F}}+ \epsilon_{_{\rm B}} \epsilon_{_{\rm F}}}{P_{_{\rm F}}}} \nu(0, \mu_6)    
        - e^{-\eta_{_{\rm F}} - \eta_{_{\rm B}}(1+\epsilon_{_{\rm F}}) }
        - \frac{e^{-\eta_{_{\rm F}}} (1 - e^{-\eta_{_{\rm B}} -\epsilon_{_{\rm B}}\eta_{_{\rm F}} }) }{1+P_{_{\rm B}}\eta_{_{\rm F}}}.
    \end{eqnarray}
\end{cor}

{\textbf{\emph{Remark 6:}}} For the case of the single GF user, the above expression for the outage probability also holds for all the feasible values of $\epsilon_{_{\rm B}}$ and $\epsilon_{_{\rm F}}$. As a result, the single GF user does not experience an outage floor when $\epsilon_{_{\rm B}} \epsilon_{_{\rm F}} \ge 1$, which will be verified by the simulation results in Section V.

\begin{theo}
    Assume that $K \ge 2$ and $P_{_{\rm B}} = P_{_{\rm F}} \to \infty$, the outage probability experienced by the admitted GF user can be approximated at high SNR as follows:
    \begin{eqnarray}
        P_{\rm out}   &\!\!\!\approx\!\!\!& \frac{\varphi_0 \epsilon_{_{\rm B}} (1+\epsilon_{_{\rm B}})^K}{P_{_{\rm F}}^{K+1}K(K-1)}   \sum\limits_{\ell = 0}^K \binom{K}{\ell} \frac{(-1)^\ell}{\ell+1}    \left( (1+\epsilon_{_{\rm F}})^{K + 1}  -  (1+\epsilon_{_{\rm F}})^{K-\ell}  \right) \nonumber \\
        &\!\!\! \!\!\!& +  \frac{\varphi_k \epsilon_{_{\rm B}}  (1+\epsilon_{_{\rm B}})^{K-k} (-1)^k }{ P_{_{\rm F}}^{K+1} } \sum\limits_{n=0}^{K-k} \binom{K-k}{n} (-1)^n (1 + \epsilon_{_{\rm F}} )^{K-k-n} \nonumber \\
    &\!\!\! \!\!\!&    \times  \sum\limits_{\ell=0}^{k} \binom{k}{\ell} (-1)^\ell     
    \frac{(1 + \epsilon_{_{\rm F}} )^{n+\ell+1} - 1}{n+\ell+1} + 
    \frac{\varphi_0 \epsilon_{_{\rm B}} \epsilon_{_{\rm F}}^K (1 + \epsilon_{_{\rm B}})(1+ \epsilon_{_{\rm F}})  }{ P_{_{\rm F}}^{K+1}K(K-1)}  
    \nonumber \\
    &\!\!\! \!\!\!&  -   
    \frac{ \varphi_0  \epsilon_{_{\rm F}}^K (\epsilon_{_{\rm B}}^{-1} + 1) (K(1+\epsilon_{_{\rm F}}) + 1) }{P_{_{\rm F}}^{K+1}K(K-1)(K+1)}     
    +   \frac{\epsilon_{_{\rm B}} \epsilon_{_{\rm F}}^{K+1} }{(K+1)P_{_{\rm F}}^{K+1} }  + \frac{\epsilon_{_{\rm F}}^K}{P_{_{\rm F}}^K}  - \frac{\epsilon_{_{\rm B}} \epsilon_{_{\rm F}}^K(1+\epsilon_{_{\rm F}})}{P_{_{\rm F}}^{K+1}} 
        \nonumber \\
        &\!\!\! \!\!\!&   +   \frac{\epsilon_{_{\rm F}}^K \left( (1 + \epsilon_{_{\rm B}})^{K+1} -1 \right) }{P_{_{\rm F}}^{K+1} (K+1)}  - \frac{\epsilon_{_{\rm F}}^K \left( (\epsilon_{_{\rm B}}(K+1) -1 )(1+\epsilon_{_{\rm B}})^{K+1} +1  \right) }{P_{_{\rm F}}^{K+2} (K+2) (K+1)}.
    \end{eqnarray}  
\end{theo}
\begin{IEEEproof}
    See Appendix B.
\end{IEEEproof}

From the results in Theorem 2, we can see that there is one term in $ P_{\rm out} $ being  proportional to $\frac{1}{P_{_{\rm F}}^K}$, while the other terms are proportional to $\frac{1}{P_{_{\rm F}}^{K+1}}$ or $\frac{1}{P_{_{\rm F}}^{K+2}}$. Therefore, we have the following corollary. 

\begin{cor}
    Assuming that $K \ge 2$ and $P_{_{\rm F}} = P_{_{\rm B}} \to \infty$, the outage probability experienced by the GF user can be further approximated as $\frac{\epsilon_{_{\rm F}}^K}{P_{_{\rm F}}^K}$ and a diversity gain of $K$ is achievable for the proposed RSMA-SGF scheme. 
\end{cor}

As a result, the RSMA-SGF scheme can achieve the diversity gain $K$ irrespective of $\epsilon_{_{\rm B}} \epsilon_{_{\rm F}} \ge 1$. 

\begin{cor}
    Assume that $K=1$, the outage probability experienced by the single GF user can be approximated as follows:
    \begin{eqnarray}
        P_{\rm out}  &\!\!\!\approx \!\!\!&  \epsilon_{_{\rm F}}P_{_{\rm F}}^{-1}.
    \end{eqnarray}
\end{cor}
\begin{IEEEproof}
    See Appendix C.
\end{IEEEproof}
 
{\textbf{\emph{Remark 7:}}} Corollaries 2 and 3 demonstrate that the proposed RSMA-SGF scheme not only avoids an outage floor, but also ensures an achievable diversity gain proportional to the number of the GF users who participate in the distributed contention. In addition, the diversity gain is achievable irrespective of the values of $\epsilon_{_{\rm B}}$ and  $\epsilon_{_{\rm F}}$, whereas the existing QoS-Guaranteed SGF (QoS-SGF) scheme requires that $\epsilon_{_{\rm B}} \epsilon_{_{\rm F}} < 1$ to achieve the diversity gain \cite{SGF_NOMA_QoS}. Thus, compared to the QoS-SGF scheme,  the RSMA-SGF scheme provides a wide ranges of target rates to achieve the diversity gain. 

\begin{cor}
    The RSMA-SGF scheme avoids an outage floor without restricting the target rates to a limited value region. 
\end{cor}
\begin{IEEEproof}
    See Appendix D. 
\end{IEEEproof}

\section{Simulation Results}

In this section, we present computer simulation results to verify the accuracy of the developed analytical results and clarify the outage performance achieved by the proposed RSMA-SGF scheme. For the purpose of comparison, two existing NOMA-SGF schemes, namely, the QoS-SGF scheme \cite{SGF_NOMA_QoS} and power control-aided SGF (denoted by PC-SGF) scheme \cite{SGF_NOMA_Advanced}, are chosen as the benchmarking schemes. It is noted that the hybrid SIC decoding order was applied in both 
the QoS-SGF and PC-SGF schemes to enhance the outage performance \cite{SGF_NOMA_QoS,SGF_NOMA_Advanced}. In addition, the optimized power control was adopted in the PC-SGF scheme \cite{SGF_NOMA_Advanced}. In the simulation, the unit for measuring the transmission data rate and target rate is bits per channel use (BPCU). 

In Fig. 2, the outage performance achieved by the proposed RSMA-SGF scheme is compared with the QoS-SGF and PC-SGF schemes for various choices of the target rate pair \{$\hat R_{_{\rm B}}$, $\hat R_{_{\rm F}}$\}. In particular, we set $P_{_{\rm F}} = \frac{P_{_{\rm B}}}{10}$ in Fig. 2 to reflect the scenarios in which the channel conditions of the GF users are weaker than that of the GB user. In Figs. 2(a) and 2(b), we set the target rate pairs as \{$\hat R_{_{\rm B}} = 1.5$ BPCU, $\hat R_{_{\rm F}} = 2$ BPCU\}  and  \{$\hat R_{_{\rm B}} = 2$ BPCU, $\hat R_{_{\rm F}} = 1.5$ BPCU\}, respectively, to indicate that the GB user has a larger or smaller target rate than that of the GF users. From Figs. 2(a) and 2(b), we can see that the proposed RSMA-SGF scheme achieves the best outage performance among three schemes for both $K = 1$ and $K = 5$. In addition, we can see that the QoS-SGF scheme causes the floors on the outage probability. The reason for this phenomenon is that QoS-Scheme requires $\epsilon_{_{\rm B}} \epsilon_{_{\rm F}} < 1$ to achieve its superior outage performance, whereas we set $\epsilon_{_{\rm B}} \epsilon_{_{\rm F}} > 1$ in Figs. 2(a) and  2(b). In fact, as indicated by \eqref{eq:eta_B_constraint}, the superior outage performance of the proposed RSMA-SGF scheme is achievable without any constraints on the values of $\epsilon_{_{\rm B}}$ and $\epsilon_{_{\rm F}}$ by avoiding the outage floor. 

In Fig. 3, we investigate the outage performance achieved by the considered SGF schemes for various transmit power settings. In particular, we assume $P_{_{\rm B}} = P_{_{\rm F}}$ in Fig. 3(a) and a fixed transmit power $P_{_{\rm B}} = 10$ dB in Fig. 3(b), respectively. The curves in Fig. 3(a) also show that the proposed RSMA-SGF scheme achieves the best outage performance without suffering outage floors. In the small and middle transmit SNR regions, the proposed scheme achieves the smaller outage probabilities than those of the PC-SGF scheme. In addition, we can see from Fig. 3(b) that when $P_{_{\rm B}}$ is fixed at 10 dB, the proposed scheme lowers the outage probability than the QoS-SGF and PC-SGF schemes 
in the whole transmit SNR region.

\begin{figure}[tb]    
    \hspace{-0.1in}
    \subfigure[$\hat R_{_{\rm B}} = 1.5$ BPCU and $\hat R_{_{\rm F}} = 2$ BPCU]{
    \includegraphics[width=3.4in]{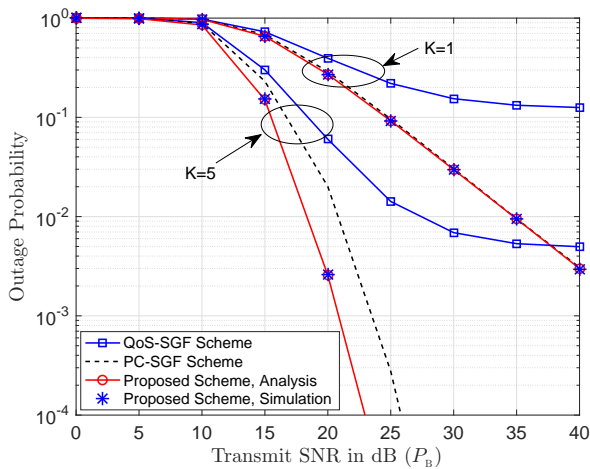}}
    \label{fig:subfig1a} \hspace{-0.2in}
    \subfigure[$\hat R_{_{\rm B}} = 2$ BPCU and $\hat R_{_{\rm F}} = 1.5$ BPCU]{
    \includegraphics[width=3.4in]{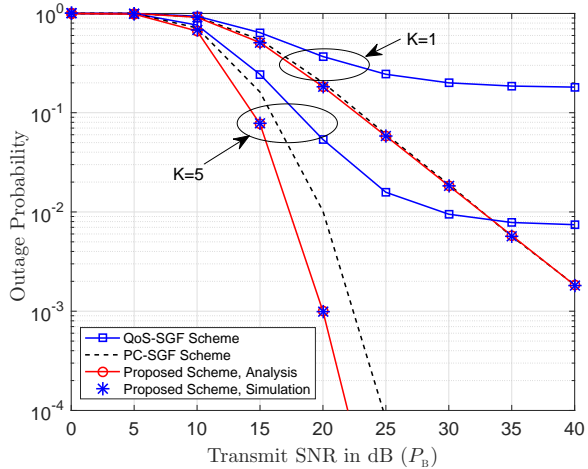}}
    \caption{Outage probability comparison of the SGF schemes for various target rate settings.}
    \label{fig:subfig1b}    
    \vspace{-0.15in}
\end{figure}

\begin{figure}[tb]    
    \hspace{-0.1in}
    \subfigure[$P_{_{\rm B}} = P_{_{\rm F}}$ ($\hat R_{_{\rm B}} = 2$ BPCU and $\hat R_{_{\rm F}} = 1.5$ BPCU)]{
    \includegraphics[width=3.4in]{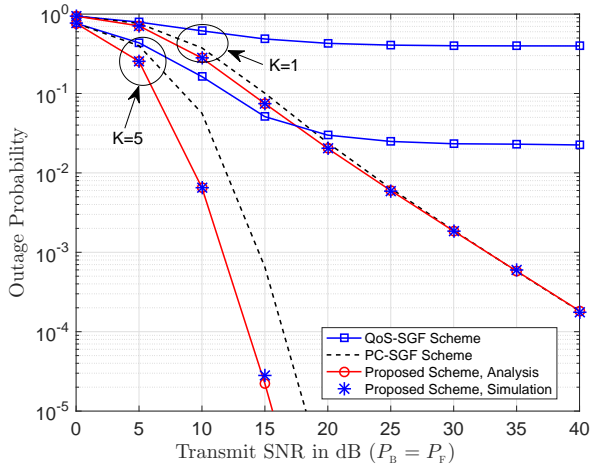}}
    \label{fig:subfig2a} \hspace{-0.2in}
    \subfigure[Fixed $P_{_{\rm B}} = 10$ dB ($\hat R_{_{\rm B}} = 1.5$ BPCU and $\hat R_{_{\rm F}} = 2$ BPCU)]{
    \includegraphics[width=3.4in]{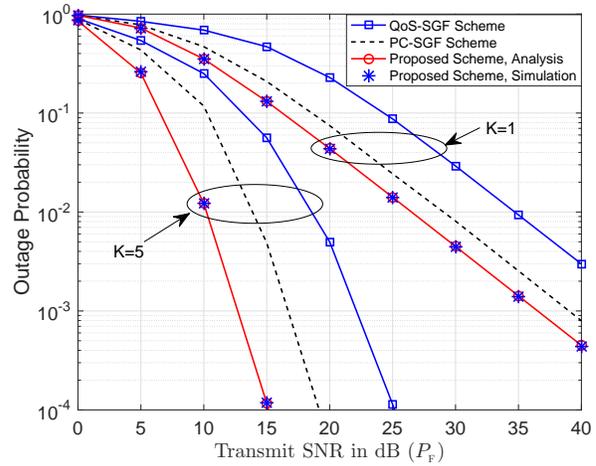}}
    \caption{Outage probability comparison of the SGF schemes for various transmit power settings.}
    \label{fig:subfig2b}    
    \vspace{-0.15in}
\end{figure}

\begin{figure}[tb]    
    \hspace{-0.1in}
    \subfigure[Exact analytical results]{
    \includegraphics[width=3.4in]{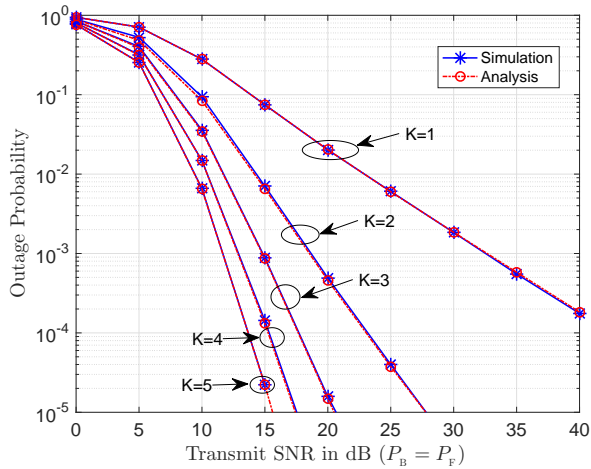}}
    \label{fig:subfig3a} \hspace{-0.2in}
    \subfigure[Approximation results]{
    \includegraphics[width=3.4in]{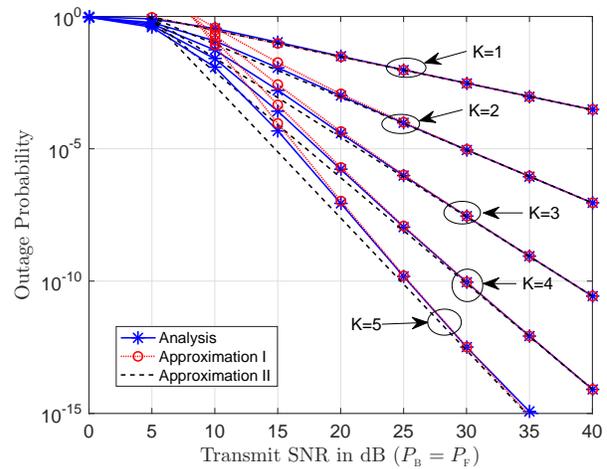}}
    \caption{Accuracy of the developed analytical results. The curves for Analysis are based Theorem 1 and Corollary 1, the curves for Approximation I are based on Theorem 2, and the curves for Approximation II are based on Corollary 2 and Corollary 3.}
    \label{fig:subfig3b}    
    \vspace{-0.15in}
\end{figure}

In Fig. 4(a), we examine the accuracy of the developed analytical results for the outage probability. In Fig. 4(a), we set \{$\hat R_{_{\rm B}} = 2$ BPCU, $\hat R_{_{\rm F}} = 1.5$ BPCU\} and use the expressions provided by Theorem 1 and Corollary 1 to obtain the analytical results. The curves in Fig. 4(a) show that the developed analytical results match the simulation results perfectly verifying the accuracy of the expressions in Theorem 1 and Corollary 1. In Fig. 4(b), we set \{$\hat R_{_{\rm B}} = 1.5$ BPCU, $\hat R_{_{\rm F}} = 2$ BPCU\} and use the developed expression in Theorem 2 to obtain the results for the curves of ``Approximation I''. In addition, the developed expressions in Corollary 2 and Corollary 3 are applied to obtain the results for the curves of ``Approximation II''. From Fig. 4(b), we can see that the curves of ``Approximation I'' match well with the analytical results in the high SNR region. Furthermore, we can see that the curves of ``Approximation I'' match well with the analytical results except for the larger values of $K$. Specifically, as the values of $K$ increase, the gaps between the curves of ``Approximation I'' and analytical results become larger since we neglect two probability related terms, $Q_k$ ($0 \le k \le K-1$) and $Q_{K+1}$, in the approximation made for Corollary 2.

\begin{figure}[tb]
    \hspace{-0.1in}    
    \subfigure[Fixed target  rate $\hat R_{_{\rm B}} = 2$ BPCU, $P_{_{\rm B}} = P_{_{\rm F}} = 15$ dB]{
    \includegraphics[width=3.4in]{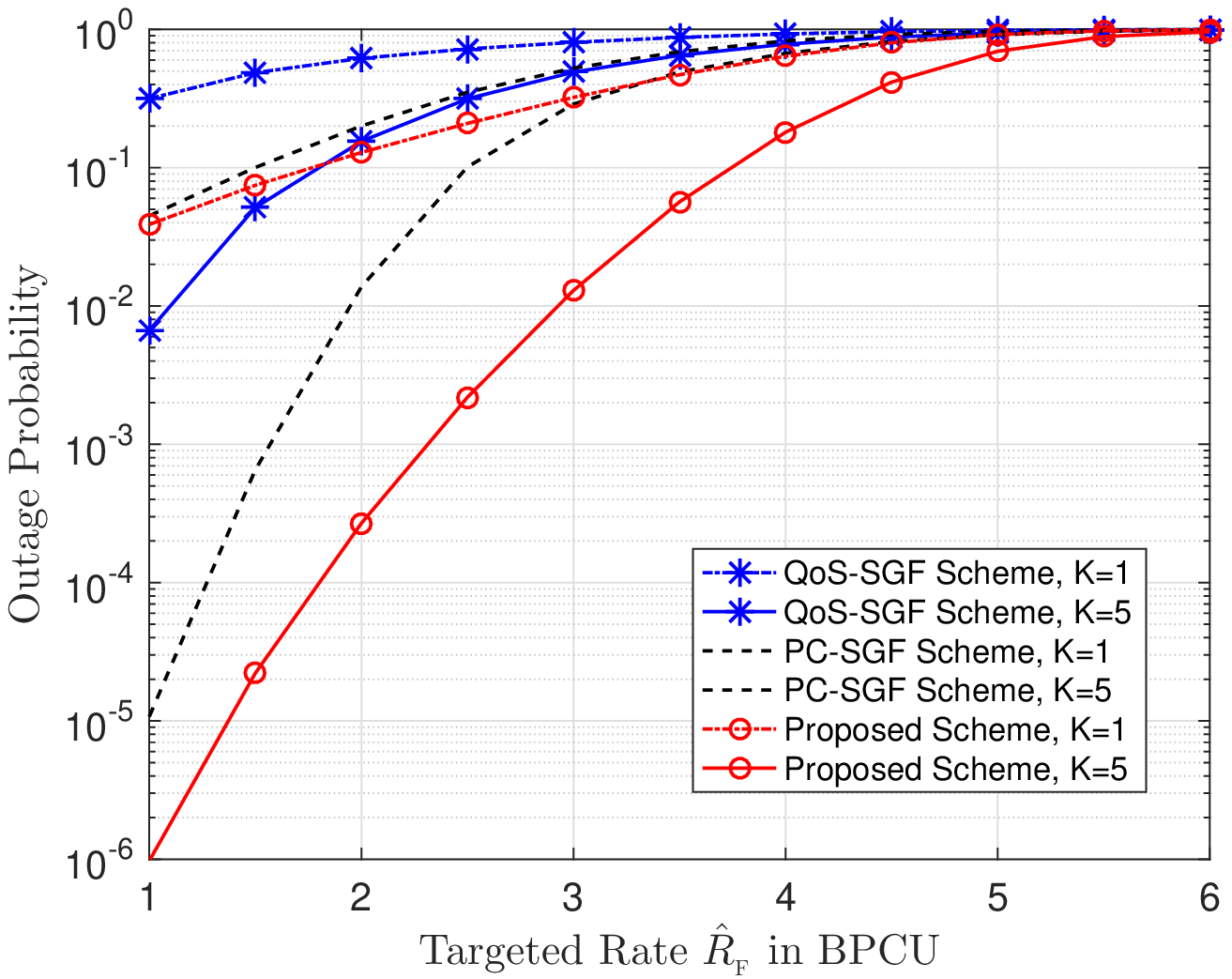}}
    \label{fig:subfig4a} \hspace{-0.2in}
    \subfigure[$P_{_{\rm B}} = P_{_{\rm F}} = 20$ dB]{
    \includegraphics[width=3.4in]{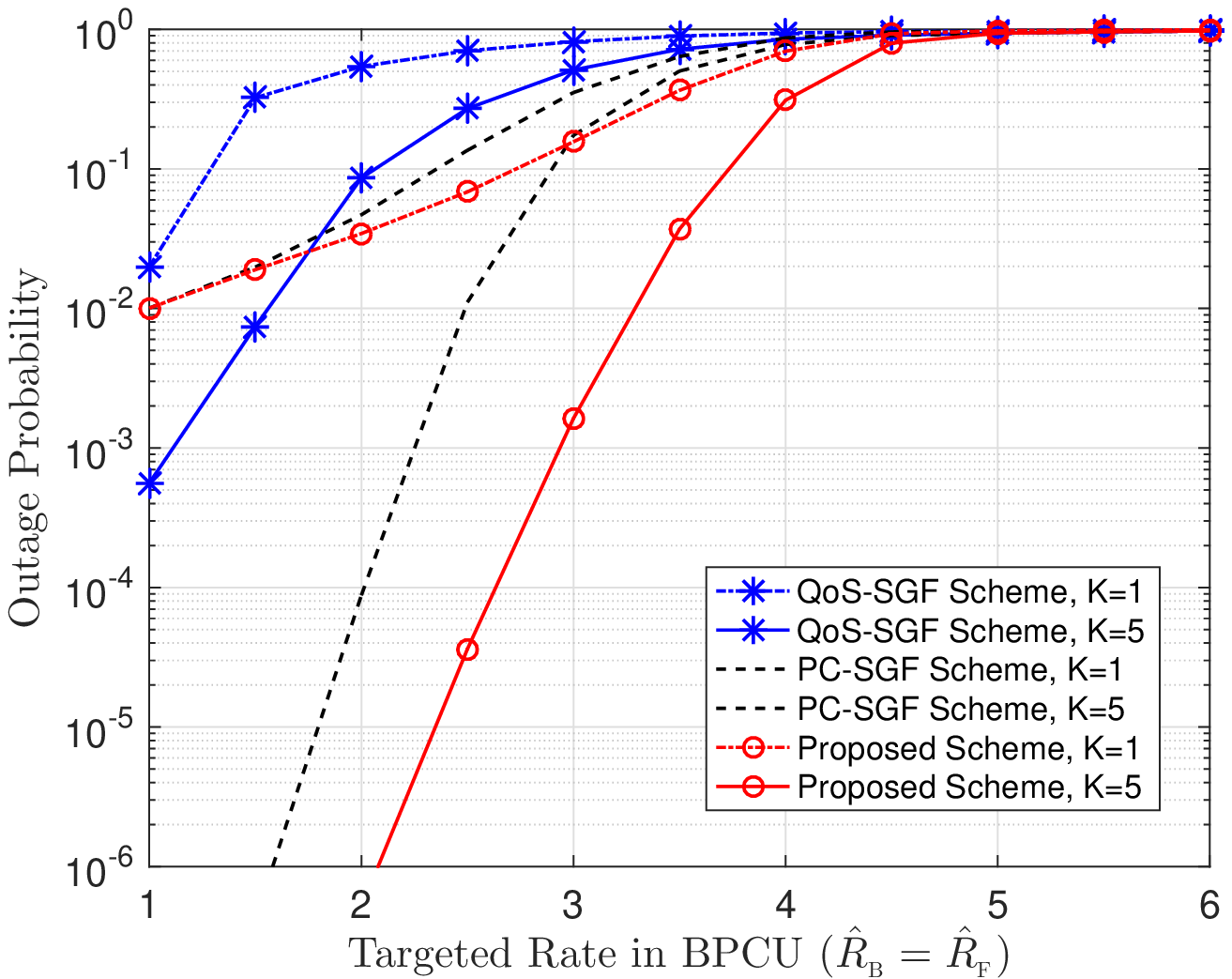}}
    \caption{Impact of the target rate on the outage probability.}
    \label{fig:subfig4b}    
    \vspace{-0.15in}
\end{figure}

\begin{figure}[tb]
    \begin{center}
    \includegraphics[width=3.4in]{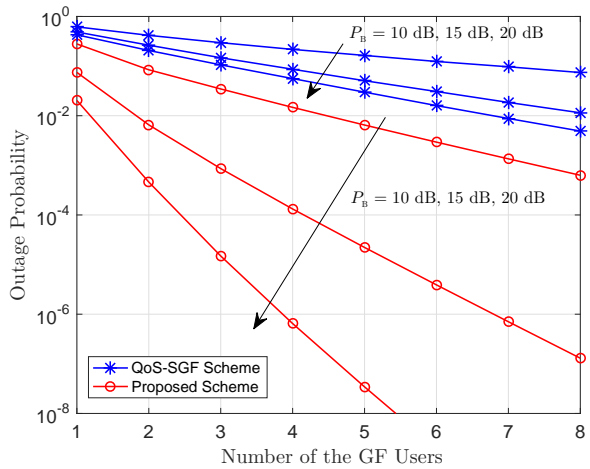}
    \caption{Impact of the number of the GF users on the outage probability ($\hat R_{_{\rm B}} = 2$ BPCU and $\hat R_{_{\rm F}} = 1.5$ BPCU, and $P_{_{\rm B}} = P_{_{\rm F}} $).}
    \label{fig:subfig3c}
    \end{center}
    \vspace{-0.15in}
\end{figure}

The impact of the target  rate on the outage probability is investigated in Fig. 5. In particular, we fix $\hat R_{_{\rm B}} = 2$ BPCU for the GB user in Fig. 5(a) and set $\hat R_{_{\rm B}} = \hat R_{_{\rm F}}$ in Fig. 5(b), respectively. It is shown in Figs. 5(a) and 5(b) that for the given transmit SNR values, the proposed RSMA-SGF scheme achieves the smallest outage probabilities in the considered whole target  rate region. As the target  rate increases, the outage probability values achieved by all the SGF schemes increase and approach 1 as the target rate increases.

The impact of the number of the GF users on the outage probability is investigated in Fig. 6, for which we set \{$\hat R_{_{\rm B}} = 2$ BPCU, $\hat R_{_{\rm F}} = 1.5$ BPCU\} and $P_{_{\rm B}} = P_{_{\rm F}}$. The results achieved by the QoS-SGF scheme are also provided for comparison. From Fig. 6, the superior outage performance achieved by the RSMA-SGF scheme over that of the QoS-SGF scheme is verified. As $K$ increases, the proposed RSMA-SGF scheme 
significantly lowers the outage probability owing to multiuser diversity. 

\begin{figure}[tb]
    \begin{center}
    \includegraphics[width=3.4in]{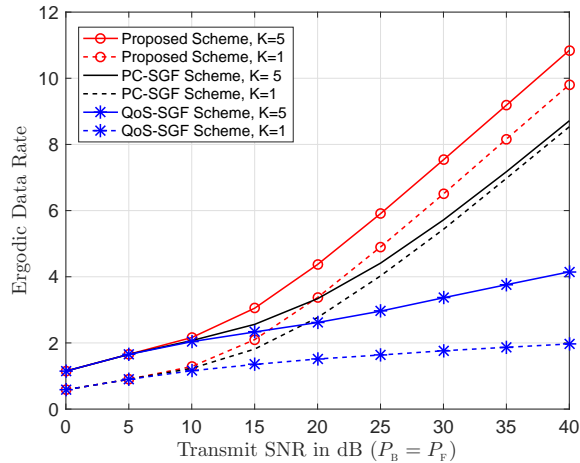}
    \caption{Ergodic data rate comparison of the SGF schemes ($\hat R_{_{\rm B}} = 4$ BPCU).}
    \label{fig:subfig3c}
    \end{center}
    \vspace{-0.15in}
\end{figure}

In Fig. 7, we evaluate the exact ergodic rate of the considered SGF schemes to verify the superior performance of the proposed RSMA-SGF scheme. It is shown in Fig. 7 that the RSMA-SGF scheme achieves the highest ergodic rate among three SGF schemes. Compared to the QoS-SGF scheme, the curve slope of the RSMA-SGF scheme is much larger, which verifies that RSMA-SGF scheme can exploit multiuser diversity more effectively than the QoS-SGF scheme. In addition, the RSMA-SGF scheme with $K=1$ even achieves a higher ergodic rate than that of the PC-SGF scheme with $K=5$ in the high SNR region, which verifies that the RSMA-SGF scheme outperforms the PC-SGF scheme. For considered high target rate $\hat R_{_{\rm B}} = 4$ BPCU, the ergodic rate gap of the PC-SGF scheme between $K=1$  and $K=5$ decreases as the SNR increases, whereas the ergodic rate gap of the RSMA-SGF scheme 
between $K=1$ and $K=5$ is almost constant in the high SNR region, which again verifies the superiority of the RSMA-SGF scheme over the existing NOMA-SGF schemes.   
 
\section{Conclusions}

In this paper, we have proposed an RSMA-SGF scheme to improve the outage performance of SGF transmissions. By applying RS, the RSMA-SGF scheme can effectively use the transmit power to exploit the advantages of power-domain superposition. Without introducing additional interference to the GB user, the proposed RSMA-SGF scheme has significantly improved the reliability of the GF users' transmissions compared to the existing NOMA-SGF schemes. To evaluate the outage performance of the RSMA-SGF scheme, we have derived exact and approximated expressions for the outage probability of the admitted GF user, which revealed that the full multiuser diversity gain is achievable irrespective of  the target rate values. Computer simulation results have been provided to clarify the superior outage performance of the proposed RSMA-SGF scheme. As one of the future research topics, user scheduling strategies considering stochastic geometry can be applied to enhance the user fairness of the RSMA-SGF scheme.     

\section*{Appendix A: A proof of Theorem 1}
\renewcommand{\theequation}{A.\arabic{equation}}
\setcounter{equation}{0}

 Since the probability term $Q_k$ involves different order statistics for different $k$ values, we evaluate different $Q_k$s considering their associated order statistics, respectively.  

 \subsection{Evaluation of $Q_0$}

The probability term $Q_0$ can be rewritten as
\begin{eqnarray}
    Q_0 &\!\!\!=\!\!\!& \Pr \left( |h_{_{\rm B}}|^2 > \eta_{_{\rm B}}, |h_1|^2 > \frac{P_{_{\rm B}}\epsilon_{_{\rm B}}^{-1}|h_{_{\rm B}}|^2 - 1 }{P_{_{\rm F}}}, |h_K|^2 < \frac{(1+\epsilon_{_{\rm F}})(1+\epsilon_{_{\rm B}})-(1 + P_{_{\rm B}}|h_{_{\rm B}}|^2 ) }{P_{_{\rm F}}}  \right) \nonumber \\
    &\!\!\!=\!\!\!& 
    \mathop{\mathcal{E}}\limits_{ \eta_{_{\rm B}}  <   |h_{_{\rm B}}|^2 < \eta_{_{\rm B}}(1+\epsilon_{_{\rm F}}) + \frac{\epsilon_{_{\rm F}}}{P_{_{\rm B}}}   }  \left\{ S_0 \right\}   \label{ap:Q_0}
\end{eqnarray} 
where $\mathop{\mathcal{E}}\{\cdot \}$ stands for the expectation operation  and $S_0$ is defined by 
\begin{eqnarray}
S_0 \triangleq \Pr \left( |h_1|^2 > \frac{P_{_{\rm B}}\epsilon_{_{\rm B}}^{-1}|h_{_{\rm B}}|^2 - 1 }{P_{_{\rm F}}}, |h_K|^2 < \frac{(1+\epsilon_{_{\rm F}})(1+\epsilon_{_{\rm B}})-(1 + P_{_{\rm B}}|h_{_{\rm B}}|^2 ) }{P_{_{\rm F}}}  \right). 
\end{eqnarray} 
In \eqref{ap:Q_0}, the expectation for $S_0$ is taken over $|h_{_{\rm B}}|^2 < \eta_{_{\rm B}}  (1+\epsilon_{_{\rm F}}) + \frac{\epsilon_{_{\rm F}}}{P_{_{\rm B}}}$ besides $|h_{_{\rm B}}|^2 > \eta_{_{\rm B}}$ due to the constraint $ \frac{(1+\epsilon_{_{\rm F}})(1+\epsilon_{_{\rm B}})-(1 + P_{_{\rm B}}|h_{_{\rm B}}|^2 ) }{P_{_{\rm F}}}$  $> 0 $, which is required by the fact that  $|h_K|^2 $ is positive in practice. 
Furthermore, since the upper bound on $|h_K|^2$ should be larger than the lower bound on $|h_1|^2$  in $S_0$,  the expectation for $S_0$ should be taken under the following  hidden constraint 
\begin{eqnarray}
    |h_{_{\rm B}}|^2 < \eta_{_{\rm B}} (1 + \epsilon_{_{\rm F}}) ,  \label{ap:hb_le_constraint}
\end{eqnarray}
which is equivalent to $\frac{(1+\epsilon_{_{\rm F}})(1+\epsilon_{_{\rm B}})-(1 + P_{_{\rm B}}|h_{_{\rm B}}|^2 ) }{P_{_{\rm F}}} > \frac{P_{_{\rm B}}\epsilon_{_{\rm B}}^{-1}|h_{_{\rm B}}|^2 - 1 }{P_{_{\rm F}}}$. Consequently, the expectation for $S_0$ should be taken over $ \eta_{_{\rm B}}  <   |h_{_{\rm B}}|^2 < \eta_{_{\rm B}}(1+\epsilon_{_{\rm F}})$ to obtain $Q_0$. 

Note that the joint probability density function (PDF) of the order statistics  $|h_1|^2$ and $|h_K|^2$ is given by
\begin{eqnarray}
    f_{|h_1|^2, |h_K|^2}(x,y) = \varphi_0 e^{-x} (e^{-x}-e^{-y})^{K-2} e^{-y}      
\end{eqnarray}
with $x < y$ and $\varphi_0 = \frac{K!}{(K-2)!}$, $S_0$ can be evaluated as follows:
\begin{eqnarray}
    S_0  &\!\!\!=\!\!\!& \varphi_0  \sum\limits_{i=0}^{K-2}  \binom{K\!-\!2}{i}  (-1)^i \int\nolimits_{\frac{\eta_{_{\rm B}}^{-1}|h_{_{\rm B}}|^2 - 1 }{P_{_{\rm F}}}}^{\frac{(1+\epsilon_{_{\rm F}})(1+\epsilon_{_{\rm B}})-(1 + P_{_{\rm B}}|h_{_{\rm B}}|^2 ) }{P_{_{\rm F}}}}  e^{-(K-i-1)x} \nonumber \\
    &\!\!\! \!\!\!& \times   \int\nolimits_{x}^{\frac{(1+\epsilon_{_{\rm F}})(1+\epsilon_{_{\rm B}})-(1 + P_{_{\rm B}}|h_{_{\rm B}}|^2 ) }{P_{_{\rm F}}}}  e^{-(i+1)y}  dydx \nonumber \\
    &\!\!\!=\!\!\!& \varphi_0  \sum\limits_{i=0}^{K-2}  \binom{K\!-\!2}{i}  \frac{(-1)^i}{i + 1}  \nonumber \\
    &\!\!\! \!\!\!& \times \left( \frac{\tilde\mu_3 e^{-\tilde\mu_4 |h_{_{\rm B}}|^2} -  \tilde\mu_5 e^{-\tilde\mu_{6} |h_{_{\rm B}}|^2}}{K}   - \frac{\tilde\mu_{1} e^{-\tilde\mu_{2} |h_{_{\rm B}}|^2} -  \tilde\mu_5 e^{-\tilde\mu_{6} |h_{_{\rm B}}|^2}}{K - i - 1}    \right), \label{ap:S0}
\end{eqnarray}
where $\tilde \mu_{1} = e^{\frac{K - (1+i)(1+\epsilon_{_{\rm B}})(1+\epsilon_{_{\rm F}})}{P_{_{\rm F}}}}$, and $\tilde \mu_{2} = \frac{K-i-1}{P_{_{\rm F}} \eta_{_{\rm B}}} - \frac{P_{_{\rm B}}(1+i)}{P_{_{\rm F}}}$, 
$\tilde \mu_3 = e^{\frac{K}{P_{_{\rm F}}}}$, $\tilde \mu_4 = \frac{K}{P_{_{\rm F}} \eta_{_{\rm B}}}$, $\tilde \mu_5 = e^{-\frac{K(\eta_{_{\rm B}} + \eta_{_{\rm F}} + \eta_{_{\rm B}} \eta_{_{\rm F}})}{P_{_{\rm F}}}}$, and $\tilde \mu_{6} = - \frac{KP_{_{\rm B}}}{P_{_{\rm F}}}$.

In order to facilitate expressing $Q_k$ ($0 \le k \le K$), we introduce a general expectation term $\nu(i, \mu)$ as follows:
\begin{eqnarray}
    \nu(i, \mu)  &\!\!\!\triangleq \!\!\!&  \mathop{\mathcal{E}}\limits_{\eta_{_{\rm B}} < |h_{_{\rm B}}|^2 < \eta_{_{\rm B}}  (1+\epsilon_{_{\rm F}})   } \left\{  e^{-\left(\frac{i  }{P_{_{\rm F}}\eta_{_{\rm B}}}  + \mu \right) |h_{_{\rm B}}|^2 } \right\}  \nonumber \\
    &\!\!\!=\!\!\!& \int\nolimits_{\eta_{_{\rm B}} }^{\eta_{_{\rm B}} (1+\epsilon_{_{\rm F}}) }  e^{-\left(\frac{i  }{P_{_{\rm F}}\eta_{_{\rm B}}}  + \mu + 1\right) x } dx  \nonumber \\
    &\!\!\!=\!\!\!&   
   \left\{ {\begin{array}{*{20}{c}}
        {\epsilon_{_{\rm F}}  \eta_{_{\rm B}}  }, &{ {\rm if~~} \mu = -1 - \frac{i}{P_{_{\rm F}}\eta_{_{\rm B}}} },\\
        {\frac{e^{- \eta_{_{\rm B}} \left(\frac{i  }{P_{_{\rm F}}\eta_{_{\rm B}}}  + \mu + 1\right) } - e^{- \eta_{_{\rm B}} (1+\epsilon_{_{\rm F}})  \left(\frac{i  }{P_{_{\rm F}}\eta_{_{\rm B}}}  + \mu + 1\right)  }}{  \frac{i }{P_{_{\rm F}}\eta_{_{\rm B}}}  + \mu + 1  } }, &{\rm otherwise}.
        \end{array}} \right.    \label{ap:nu_term}
\end{eqnarray}  
Applying the results of \eqref{ap:nu_term} into \eqref{ap:S0}, $Q_0$ can be calculated as
\begin{eqnarray}
    Q_0 &\!\!\!=\!\!\!& \varphi_0  \sum\limits_{i=0}^{K-2}  \binom{K\!-\!2}{i}  \frac{(-1)^i}{i + 1}  \left( \frac{\tilde \mu_3 \nu(0, \tilde \mu_4) -  \tilde \mu_5 \nu(0,\tilde \mu_{6})}{K}   - \frac{\tilde \mu_{1} \nu(0,\tilde \mu_{2})  -  \tilde \mu_5 \nu(0,\tilde \mu_{6}) }{K - i - 1}    \right)\!\!. ~~~~~~~ \label{ap:Q_0_2}
\end{eqnarray}
Since that $Q_0$ in \eqref{ap:Q_0_2} is too complicated to obtain a high SNR approximation, we simplify it in the following way.

By applying $\binom{K-2}{i} = \binom{K-1}{i+1} \frac{i+1}{K-1}$ and substituting $\ell = i+ 1$ into \eqref{ap:Q_0_2}, $Q_0$ can be rewritten as follows:
\begin{eqnarray}
    Q_0 &\!\!\!=\!\!\!&\! -\frac{\varphi_0}{K\!-\!1} \! \sum\limits_{\ell=0}^{K-1} \! \binom{K\!-\!1}{\ell}  \!(-1)^\ell \! \left( \frac{\tilde \mu_3 \nu(0, \tilde \mu_4) \!-\!  \tilde \mu_5 \nu(0, \tilde \mu_{6})}{K}   \!-\! \frac{\mu_{1} \nu(0,\mu_{2})  \!-\!  \tilde \mu_5 \nu(0,\tilde \mu_{6}) }{K - \ell}    \right)\!\!, ~~~~~~~ \label{ap:Q_0_3}
\end{eqnarray}
where $\mu_{1} = e^{\frac{K - \ell(1+\epsilon_{_{\rm B}})(1+\epsilon_{_{\rm F}})}{P_{_{\rm F}}}}$ and $\mu_{2} = \frac{K-\ell}{P_{_{\rm F}} \eta_{_{\rm B}}} - \frac{P_{_{\rm B}}\ell}{P_{_{\rm F}}}$. In \eqref{ap:Q_0_3}, we add the term for $\ell=0$ without changing the summation value due to the fact that $\frac{\tilde \mu_3 \nu(0, \tilde \mu_4) \!-\!  \tilde \mu_5 \nu(0, \tilde \mu_{6})}{K}   \!-\! \frac{\mu_{1} \nu(0,\mu_{2})  \!-\!  \tilde \mu_5 \nu(0,\tilde \mu_{6}) }{K - \ell} =0$ when $\ell = 0$. 
By eliminating the terms that are independent of $\ell$ based on the fact that $\sum\nolimits_{\ell=0}^{K-1}  \binom{K-1}{\ell}  (-1)^\ell = 0$, $Q_0$ can be simplified as:
\begin{eqnarray}
    Q_0 &\!\!\!=\!\!\!& \frac{\varphi_0}{K-1}  \sum\limits_{\ell=0}^{K-1}  \binom{K\!-\!1}{\ell}   (-1)^\ell  ~ \frac{\mu_{1} \nu(0, \mu_{2})  -  \tilde \mu_5 \nu(0,\tilde\mu_{6}) }{K - \ell}  \nonumber \\
    &\!\!\!=\!\!\!&   \frac{\varphi_0}{K(K-1)}  \sum\limits_{\ell=0}^{K}  \binom{K}{\ell}   (-1)^\ell  \left( \mu_{1} \nu(0,  \mu_{2})  -  \tilde \mu_5 \nu(0,\tilde \mu_{6}) \right)  , \label{ap:Q_0_4}
\end{eqnarray}
where the term for $\ell = K$ is added without changing the summation value due to the fact that $ \mu_{1} \nu(0,  \mu_{2})  -  \tilde \mu_5 \nu(0,\tilde \mu_{6}) = 0$ when  $\ell = K$.  

Again, by eliminating the terms that are independent of $\ell$ based on the fact that $\sum\nolimits_{\ell=0}^{K}  \binom{K}{\ell}  (-1)^\ell = 0$, $Q_0$ can be further simplified as:
\begin{eqnarray}
    Q_0 &\!\!\!=\!\!\!& \frac{\varphi_0}{K(K-1)}  \sum\limits_{\ell=0}^{K}  \binom{K}{\ell}   (-1)^\ell  ~ \mu_{1} \nu(0, \mu_{2})  . \label{ap:Q_0_5}
\end{eqnarray}

 \subsection{Evaluation of $Q_k$ for $1 \le k \le K-2$}

 When $1 \le k \le K-2$, three order statistics, $h_k$, $h_{k+1}$, and $h_K$, are involved in probability $Q_k$ in the form of 
 \begin{eqnarray}
     Q_k &\!\!\!=\!\!\!& \Pr \left( |h_{_{\rm B}}|^2 > \eta_{_{\rm B}},  |h_k|^2 < \frac{\tau}{P_{_{\rm F}}}, |h_{k+1}|^2 > \frac{\tau}{P_{_{\rm F}}} , R_{_{{\rm II},K}}  < \hat R_{_{\rm F}} \right) \nonumber \\
     &\!\!\!=\!\!\!&   \mathop{\mathcal{E}}\limits_{\eta_{_{\rm B}} < |h_{_{\rm B}}|^2 < \eta_{_{\rm B}} (1+\epsilon_{_{\rm F}})   } \{ S_k \}, \label{ap:Q_k}
 \end{eqnarray} 
 where $S_k$ is defined by 
 \begin{eqnarray}
     S_k &\!\!\! \triangleq \!\!\!& \Pr \left(  |h_k|^2 < \frac{P_{_{\rm B}}\epsilon_{_{\rm B}}^{-1}|h_{_{\rm B}}|^2 - 1 }{P_{_{\rm F}}}, |h_{k+1}|^2 > \frac{P_{_{\rm B}}\epsilon_{_{\rm B}}^{-1}|h_{_{\rm B}}|^2 - 1 }{P_{_{\rm F}}} , \right.  \nonumber \\
     &\!\!\! \!\!\!& ~~~~~   \left.  |h_K|^2 < \frac{(1+\epsilon_{_{\rm F}})(1+\epsilon_{_{\rm B}})-(1 + P_{_{\rm B}}|h_{_{\rm B}}|^2 ) }{P_{_{\rm F}}}  \right). \label{ap:S_k_0}
 \end{eqnarray}
 In \eqref{ap:Q_k}, the expectation is taken over $\eta_{_{\rm B}} < |h_{_{\rm B}}|^2 < \eta_{_{\rm B}} (1+\epsilon_{_{\rm F}}) $ considering the relationship between the upper and lower bounds on channel gains.  
 
For the case $1 \le k \le K-2$, the joint PDF of three order statistics, $h_k$, $h_{k+1}$, and $h_K$, is given by 
\begin{eqnarray}
    f_{|h_k|^2, |h_{k+1}|^2, |h_K|^2} (x, y, z) &\!\!\!\!=\!\!\!\!& \tilde \varphi_k e^{-x} (1-e^{-x})^{k-1} e^{-y} (e^{-y} - e^{-z})^{K-k-2} e^{-z} \nonumber \\
    &\!\!\!\!=\!\!\!\!& \tilde \varphi_k \!\! \sum\limits_{i=0}^{\!K\!-k\!-\!2\!} \!\! \binom{K\!-\!k\!-\!2}{i} \! (-1)^i e^{-x} (1-e^{-x})^{k-1} e^{-(K-k-i-1)y} e^{-(i+1)z}, ~~~~~~ 
\end{eqnarray}
where $x \le y \le z$ and $\tilde \varphi_k = \frac{K!}{(k-1)! (K-k-2)!}$. Using the above joint PDF, $S_k$ can be expressed in terms of $|h_{_{\rm B}}|^2$ as follows:
\begin{eqnarray}
    S_k  &\!\!\!=\!\!\!&\tilde \varphi_k \! \sum\limits_{i=0}^{K-k-2} \!\! \binom{K\!-\!k\!-\!2}{i}  (-1)^i \int\nolimits_{0}^{\frac{\eta_{_{\rm B}}^{-1}|h_{_{\rm B}}|^2 - 1 }{P_{_{\rm F}}}} e^{-x} (1-e^{-x})^{k-1} \nonumber \\
    &\!\!\! \!\!\!& \times \int\nolimits_{\frac{\eta_{_{\rm B}}^{-1}|h_{_{\rm B}}|^2 - 1 }{P_{_{\rm F}}}}^{\frac{(1+\epsilon_{_{\rm F}})(1+\epsilon_{_{\rm B}})-(1 + P_{_{\rm B}}|h_{_{\rm B}}|^2 ) }{P_{_{\rm F}}}}  e^{-(K-k-i-1)y}  \int\nolimits_{y}^{\frac{(1+\epsilon_{_{\rm F}})(1+\epsilon_{_{\rm B}})-(1 + P_{_{\rm B}}|h_{_{\rm B}}|^2 ) }{P_{_{\rm F}}}}  e^{-(i+1)z}  dzdydx . ~~~~
\end{eqnarray}
After some algebraic manipulations, $S_k$ can be evaluated as follows:
\begin{eqnarray}
    S_k  &\!\!\!=\!\!\!& \tilde \varphi_k \! \sum\limits_{i=0}^{K-k-2} \!\! \binom{K\!-\!k\!-\!2}{i} \sum\limits_{\ell=0}^{k} \binom{k}{\ell} \frac{(-1)^{\ell+i}e^{\frac{\ell}{P_{_{\rm F}}}} e^{-\frac{\ell|h_{_{\rm B}}|^2  }{P_{_{\rm F}}\eta_{_{\rm B}}} }}{k (i+1)} \nonumber \\
    &\!\!\! \!\!\!& \times \left( \frac{\tilde \mu_1 e^{-\tilde \mu_2 |h_{_{\rm B}}|^2} -  \tilde \mu_5 e^{-\tilde \mu_6 |h_{_{\rm B}}|^2}}{K - k}   - \frac{\tilde \mu_3 e^{-\tilde \mu_4 |h_{_{\rm B}}|^2} -  \tilde \mu_5 e^{-\tilde \mu_6 |h_{_{\rm B}}|^2}}{K - k - i - 1}    \right), \label{ap:S_k}
\end{eqnarray}
where $\tilde \mu_1 = e^{\frac{K-k}{P_{_{\rm F}}}}$, $\tilde \mu_2 = \frac{K-k}{P_{_{\rm F}}\eta_{_{\rm B}}}$,  $\tilde \mu_3 = e^{\frac{K-k - (1+i) \left(1+\epsilon_{_{\rm B}}\right) \left(1+ \epsilon_{_{\rm F}}\right) }{P_{_{\rm F}}}}$, $\tilde \mu_4 = \frac{K-k-i-1}{P_{_{\rm F}} \eta_{_{\rm B}} }  - \frac{(1+i)P_{_{\rm B}}}{P_{_{\rm F}}}  $, 
$\tilde \mu_5 = e^{-\frac{(K-k)\left(\epsilon_{_{\rm B}} + \epsilon_{_{\rm F}} + \epsilon_{_{\rm B}}\epsilon_{_{\rm F}}  \right)}{P_{_{\rm F}} }}$, and $\tilde \mu_6 = -\frac{(K - k)P_{_{\rm B}} }{P_{_{\rm F}} } $. 

Taking the expectation for $S_k$ according to \eqref{ap:Q_k}, $Q_k$ can be evaluated as:
\begin{eqnarray}
    Q_k  &\!\!\!=\!\!\!& \tilde \varphi_k \! \sum\limits_{i=0}^{K-k-2} \!\! \binom{K\!-\!k\!-\!2}{i} \frac{(-1)^i }{k (i+1)} \sum\limits_{\ell=0}^{k} \binom{k}{\ell} (-1)^{\ell}e^{\frac{\ell}{P_{_{\rm F}}}} \nonumber \\
    &\!\!\! \!\!\!& \times \left( \frac{\tilde \mu_1 \nu(\ell, \tilde \mu_2)  -  \tilde \mu_5 \nu(\ell, \tilde \mu_6)}{K - k}   - \frac{\tilde \mu_3 \nu(\ell, \tilde \mu_4) -  \tilde \mu_5 \nu(\ell, \tilde \mu_6) }{K - k - i - 1}    \right), \label{ap:Q_k_3}
\end{eqnarray}
where $\nu(\ell, \tilde \mu)$ is given by \eqref{ap:nu_term}. 
For the expression in \eqref{ap:Q_k_3}, we can replace $\binom{K-k-2}{i}$ with $\binom{K-k-1}{i+1} \frac{i+1}{K-k-1}$ and let $n = i+ 1$,  so that $Q_k$ can be rewritten as:
\begin{eqnarray}
    Q_k  &\!\!\!=\!\!\!& \frac{-\tilde \varphi_k}{ k (K-k-1)} \! \sum\limits_{n=0}^{K\!-\!k\!-\!1} \!\! \binom{K-k-1}{n} (-1)^n \sum\limits_{\ell=0}^{k} \binom{k}{\ell} (-1)^{\ell}e^{\frac{\ell}{P_{_{\rm F}}}} \nonumber \\
    &\!\!\! \!\!\!& \times \left( \frac{\tilde \mu_1 \nu(\ell, \tilde \mu_2)  -  \tilde \mu_5 \nu(\ell, \tilde \mu_6)}{K - k}   - \frac{ \mu_3 \nu(\ell,  \mu_4) -  \tilde \mu_5 \nu(\ell, \tilde \mu_6) }{K - k - n}    \right), \label{ap:Q_k_4}
\end{eqnarray}
where $ \mu_3 = e^{\frac{K-k - n \left(1+\epsilon_{_{\rm B}} \right) \left(1+ \epsilon_{_{\rm F}} \right) }{P_{_{\rm F}}}}$  and $ \mu_4 = \frac{K-k-n}{P_{_{\rm F}} \eta_{_{\rm B}} }  - \frac{n P_{_{\rm B}}}{P_{_{\rm F}}}  $. For \eqref{ap:Q_k_4}, we note that the added term for $n=0$ does not change the value of $Q_k$ due to the fact that $\frac{\tilde \mu_1 \nu(\ell, \tilde \mu_2)  -  \tilde \mu_5 \nu(\ell, \tilde \mu_6)}{K - k}   - \frac{ \mu_3 \nu(\ell,  \mu_4) -  \tilde \mu_5 \nu(\ell, \tilde \mu_6) }{K - k - n} = 0$ when $n=0$. 

Furthermore, some terms in \eqref{ap:Q_k_4} invlolving $\tilde \mu_1$, $\tilde \mu_2$, $\tilde \mu_5$, and $\tilde \mu_6$ but being independent of $n$ can be further eliminated due to the fact that $\sum\nolimits_{n=0}^k \binom{k}{n} (-1)^n = 0$, while $\tilde \mu_1$, $\tilde \mu_2$, $\tilde \mu_5$, and $\tilde \mu_6$ are not functions of $n$. The simplification can be expressed as follows:
\begin{eqnarray}
    Q_k  &\!\!\!=\!\!\!& \frac{\tilde \varphi_k}{ k (K-k-1)} \! \sum\limits_{n=0}^{K-k-1} \!\! \binom{K\!-\!k\!-\!1}{n} (-1)^n \sum\limits_{\ell=0}^{k} \binom{k}{\ell} (-1)^{\ell}e^{\frac{\ell}{P_{_{\rm F}}}}   \frac{ \mu_3 \nu(\ell,  \mu_4) -  \tilde \mu_5 \nu(\ell, \tilde \mu_6) }{K - k - n}   \nonumber \\
    &\!\!\! \mathop  = \limits^{(a)} \!\!\!&   \varphi_k  \sum\limits_{n=0}^{K-k}   \binom{K\!-\!k}{n} (-1)^n \sum\limits_{\ell=0}^{k} \binom{k}{\ell} (-1)^{\ell}e^{\frac{\ell}{P_{_{\rm F}}}}  ( \mu_3 \nu(\ell, \mu_4) -  \tilde \mu_5 \nu(\ell, \tilde \mu_6))  \nonumber \\
    &\!\!\! \mathop  = \limits^{(b)} \!\!\!&  \varphi_k  \sum\limits_{n=0}^{K-k}   \binom{K\!-\!k}{n} (-1)^n \sum\limits_{\ell=0}^{k} \binom{k}{\ell} (-1)^{\ell}e^{\frac{\ell}{P_{_{\rm F}}}}  \mu_3 \nu(\ell,   \mu_4) ,    
\end{eqnarray}
where step (a) follows by absorbing $K-k-1$ into the binomial coefficients without changing the summation value and step (b) follows by using the fact that $\sum\nolimits_{n=0}^k \binom{k}{n} (-1)^n = 0$  while $\tilde \mu_5$  and $\tilde \mu_6$ are not functions of $n$. 

\subsection{Evaluation of $Q_{K-1}$}

\begin{eqnarray}
    Q_{K-1} &\!\!\!=\!\!\!&  \mathop{\mathcal{E}}\limits_{|h_{_{\rm B}}|^2 > \eta_{_{\rm B}} }  \left\{  \Pr \left( |h_{K-1}|^2 < \frac{P_{_{\rm B}}\epsilon_{_{\rm B}}^{-1}|h_{_{\rm B}}|^2 - 1 }{P_{_{\rm F}}}, |h_K|^2 > \frac{P_{_{\rm B}}\epsilon_{_{\rm B}}^{-1}|h_{_{\rm B}}|^2 - 1 }{P_{_{\rm F}}}  \right. \right.  \nonumber \\
    &\!\!\! \!\!\!& ~~~~~~~~~~~   \left. \left.  |h_K|^2 < \frac{(1+\epsilon_{_{\rm F}})(1+\epsilon_{_{\rm B}})-(1 + P_{_{\rm B}}|h_{_{\rm B}}|^2 ) }{P_{_{\rm F}}}  \right) \right\}   \label{ap:Q_K11}
\end{eqnarray} 
By extracting the hidden constraint on the upper and lower bounds on $|h_K|^2$ from \eqref{ap:Q_K11}, i.e., $ \frac{P_{_{\rm B}}\epsilon_{_{\rm B}}^{-1}|h_{_{\rm B}}|^2 - 1 }{P_{_{\rm F}}} < \frac{(1+\epsilon_{_{\rm F}})(1+\epsilon_{_{\rm B}})-(1 + P_{_{\rm B}}|h_{_{\rm B}}|^2 ) }{P_{_{\rm F}}}$, $Q_{K-1}$ can be rewritten as follows:
\begin{eqnarray}
    Q_{K-1} &\!\!\!=\!\!\!&  \mathop{\mathcal{E}}\limits_{\eta_{_{\rm B}} < |h_{_{\rm B}}|^2 < (1+\epsilon_{_{\rm F}}) \eta_{_{\rm B}}  } \{ S_{K-1} \},  \label{ap:Q_k_11}
\end{eqnarray} 
where $S_{K-1}$ denotes probability inside the expectation in \eqref{ap:Q_K11}. There are two order statistics $h_{K-1}$ and $h_{K}$ involving in $S_{K-1}$ with their joint PDF being given by 
\begin{eqnarray}
    f_{|h_{K-1}|^2, |h_K|^2} (x, y) = \varphi_0 e^{-x} (1-e^{-x})^{K-2} e^y,
\end{eqnarray}
where $x \le y$. With the aid of this joint PDF, $S_{K-1}$ can be evaluated as follows:
\begin{eqnarray}
    S_{K-1} = \varphi_0 \sum\limits_{\ell=0}^{K-1} \binom{K\!-\!1}{\ell} \frac{(-1)^\ell e^{\frac{\ell}{P_{_{\rm F}}} } e^{-\frac{\ell|h_{_{\rm B}}|^2  }{P_{_{\rm F}}\eta_{_{\rm B}}} } }{K-1}  \left( e^{\frac{1}{P_{_{\rm F}}}} e^{-\mu_5 |h_{_{\rm B}}|^2}  - e^{-\frac{\epsilon_{_{\rm B}} + \epsilon_{_{\rm F}} + \epsilon_{_{\rm B}}\epsilon_{_{\rm F}}  }{P_{_{\rm F}}} } e^{-\mu_6 |h_{_{\rm B}}|^2}    \right), ~~
\end{eqnarray}
where $\mu_5 = \frac{1}{P_{_{\rm F}}\eta_{_{\rm B}}}$ and $\mu_6 = -\frac{P_{_{\rm B}}}{P_{_{\rm F}}}$. 
By using the results in \eqref{ap:nu_term}, $Q_{K-1}$ can be derived as follows:
\begin{eqnarray}
    Q_{K-1} = \frac{\varphi_0}{K-1} \sum\limits_{\ell=0}^{K-1} \binom{K\!-\!1}{\ell}  (-1)^\ell e^{\frac{\ell}{P_{_{\rm F}}} }    \left( e^{\frac{1}{P_{_{\rm F}}}} \nu(\ell, \mu_5)  - e^{-\frac{\epsilon_{_{\rm B}} + \epsilon_{_{\rm F}} + \epsilon_{_{\rm B}}\epsilon_{_{\rm F}}  }{P_{_{\rm F}}} } \nu(\ell, \mu_6)   \right). 
\end{eqnarray}

\subsection{Evaluation of $Q_K$ and $Q_{K+1}$}

In the case of $Q_K$, the determination of the probability value involves two independent random variables $|h_{_{\rm B}}|^2$ and $|h_K|^2$. 
Recalling the expression in \eqref{eq:Pout_GF}, $Q_K$ can be rewritten as follows:
\begin{eqnarray}
    Q_K =  \mathop{\mathcal{E}}\limits_{|h_{_{\rm B}}|^2 > \eta_{_{\rm B}} }  \left\{   \Pr \left( |h_K|^2 < \frac{\eta_{_{\rm B}}^{-1}|h_{_{\rm B}}|^2 - 1 }{P_{_{\rm F}}},   |h_K|^2 < \frac{\epsilon_{_{\rm F}} }{ P_{_{\rm F}} }  \right)  \right\} .    
\end{eqnarray}
Comparing $\frac{\eta_{_{\rm B}}^{-1}|h_{_{\rm B}}|^2 - 1 }{P_{_{\rm F}}}$ and $ \frac{\epsilon_{_{\rm F}} }{ P_{_{\rm F}} }$, it can be shown that $\frac{\eta_{_{\rm B}}^{-1}|h_{_{\rm B}}|^2 - 1 }{P_{_{\rm F}}} < \frac{\epsilon_{_{\rm F}} }{ P_{_{\rm F}} }$ if $|h_{_{\rm B}}|^2  < \eta_{_{\rm B}} (1+\epsilon_{_{\rm F}})$; otherwise,  $\frac{\eta_{_{\rm B}}^{-1}|h_{_{\rm B}}|^2 - 1 }{P_{_{\rm F}}} > \frac{\epsilon_{_{\rm F}} }{ P_{_{\rm F}} }$.  Therefore, $Q_K$ can be evaluated as follows:
\begin{eqnarray}
    Q_K &\!\!\!=\!\!\!&  \mathop{\mathcal{E}}\limits_{ \eta_{_{\rm B}} < |h_{_{\rm B}}|^2  < \eta_{_{\rm B}} (1+\epsilon_{_{\rm F}}) }  \left\{   \Pr \left( |h_K|^2 < \frac{\eta_{_{\rm B}}^{-1}|h_{_{\rm B}}|^2 - 1 }{P_{_{\rm F}}} \right)  \right\}  + \mathop{\mathcal{E}}\limits_{  |h_{_{\rm B}}|^2  > \eta_{_{\rm B}} (1+\epsilon_{_{\rm F}}) }  \left\{   \Pr \left(  |h_K|^2 <  \eta_{_{\rm F}}   \right)  \right\}  \nonumber \\
    &\!\!\!= \!\!\!& \int\nolimits_{\eta_{_{\rm B}}}^{ \eta_{_{\rm B}} (1+\epsilon_{_{\rm F}})} \left(1 - e^{-\frac{\eta_{_{\rm B}}^{-1}x - 1 }{P_{_{\rm F}}} } \right)^K e^{-x} dx + \left(1 - e^{-\eta_{_{\rm F}}}  \right)^K e^{- \eta_{_{\rm B}} (1+\epsilon_{_{\rm F}})  }      \nonumber \\
    &\!\!\!= \!\!\!& \sum\limits_{\ell = 0}^K \binom{K}{\ell} (-1)^\ell e^{\frac{\ell}{P_{_{\rm F}}}} \int\nolimits_{\eta_{_{\rm B}} }^{\eta_{_{\rm B}} (1+\epsilon_{_{\rm F}}) }  e^{-\left(\frac{\ell  }{P_{_{\rm F}}\eta_{_{\rm B}}}   + 1\right) x } dx    + \left(1 - e^{-\eta_{_{\rm F}}}  \right)^K e^{- \eta_{_{\rm B}} (1+\epsilon_{_{\rm F}})  }    \nonumber \\
    &\!\!\!= \!\!\!&  \sum\limits_{\ell = 0}^K \binom{K}{\ell} (-1)^\ell e^{\frac{\ell}{P_{_{\rm F}}}} \nu(\ell, 0)   + \left(1 - e^{-\eta_{_{\rm F}}}  \right)^K e^{- \eta_{_{\rm B}} (1+\epsilon_{_{\rm F}})  }
\end{eqnarray}  

Similarly to $Q_K$, $Q_{K+1}$ is a function of two independent random variables $|h_{_{\rm B}}|^2$ and $|h_K|^2$. Recalling that $R_{_{{\rm II},K}} =  \log_2\left( 1 + \frac{P_{_{\rm F}}|h_K|^2 }{P_{_{\rm B}}|h_{_{\rm B}}|^2  + 1} \right)$ when $\tau = 0$ occurs, $Q_{K+1}$ can be evaluated as follows:
\begin{eqnarray}
    Q_{K+1} &\!\!\!= \!\!\!&  \Pr \left( |h_{_{\rm B}}|^2 < \eta_{_{\rm B}},  \log_2\left( 1 + \frac{P_{_{\rm F}}|h_K|^2 }{P_{_{\rm B}}|h_{_{\rm B}}|^2  + 1} \right) < \hat R_{_{\rm F}}  \right)  \nonumber \\ 
    &\!\!\!= \!\!\!& \sum\limits_{\ell = 0}^K \binom{K}{\ell}  (-1)^\ell e^{-\ell \eta_{_{\rm F}}} \frac{1 - e^{-(1+\ell \eta_{_{\rm F}} P_{_{\rm B}} )\eta_{_{\rm B}} }}{1 + \ell \eta_{_{\rm F}}  P_{_{\rm B}}}.
\end{eqnarray}

\section*{Appendix B: A proof of Theorem 2}
\renewcommand{\theequation}{B.\arabic{equation}}
\setcounter{equation}{0}
\setcounter{subsection}{0}

Regarding that $Q_k$ depends on the value of $k$, the high SNR approximations for different $Q_k$ will be tackled separately in the following subsections.  

\subsection{High SNR approximation for $Q_0$}

Based on the derived closed-form expression in \eqref{ap:Q_0_5}, $Q_0$ can be rewritten as follows:
\begin{eqnarray}
    Q_0 &\!\!\!=\!\!\!& \frac{\varphi_0}{K(K-1)}  \sum\limits_{\ell=0}^{K}  \binom{K}{\ell}   (-1)^\ell   \mu_{1} \nu(0, \mu_{2}) . \nonumber \\
    &\!\!\!=\!\!\!& \frac{\varphi_0}{K(K-1)}  \sum\limits_{\ell=0}^{K}  \binom{K}{\ell}   (-1)^\ell \mu_1 \int\nolimits_{\eta_{_{\rm B}}}^{\eta_{_{\rm B}}(1+\epsilon_{_{\rm F}})} e^{-(\mu_2+1)x} dx \label{apb:Q_0}
\end{eqnarray}
By applying the approximation, $e^{-x} = 1 - x$ for $x \to 0$, and using the definitions of $\mu_1$ and $\mu_2$, $Q_0$ can be approximated as follows:
\begin{eqnarray}
    Q_0 &\!\!\!=\!\!\!& \frac{\varphi_0}{K(K-1)}  \int\nolimits_{\eta_{_{\rm B}}}^{\eta_{_{\rm B}}(1+\epsilon_{_{\rm F}})} \sum\limits_{\ell=0}^{K}  \binom{K}{\ell}   (-1)^\ell  e^{-\frac{\ell (1+\epsilon_{_{\rm B}} )(1+ \epsilon_{_{\rm F}})}{P_{_{\rm F}}}}  e^{\ell (\epsilon_{_{\rm B}}^{-1}  + 1  )x}  dx   \nonumber \\
    &\!\!\! \mathop = \limits^{(a)} \!\!\!&  \frac{\varphi_0}{K(K-1)}  \int\nolimits_{\eta_{_{\rm B}}}^{\eta_{_{\rm B}}(1+\epsilon_{_{\rm F}})}  \left( 1  -  e^{-\left( \frac{ (1+\epsilon_{_{\rm B}} )(1+ \epsilon_{_{\rm F}})}{P_{_{\rm F}}} - (\epsilon_{_{\rm B}}^{-1}  +1) x  \right) } \right)^K  dx  \nonumber \\
    &\!\!\!=\!\!\!&    \frac{\varphi_0}{K(K-1)}  \int\nolimits_{\eta_{_{\rm B}}}^{\eta_{_{\rm B}}(1+\epsilon_{_{\rm F}})} \left(  \frac{ (1+\epsilon_{_{\rm B}} )(1+ \epsilon_{_{\rm F}})}{P_{_{\rm F}}} - \left(\epsilon_{_{\rm B}}^{-1}  +1 \right) x   \right)^K dx, 
\end{eqnarray}
where step (a) is obtained by using the fact that $\sum\nolimits_{\ell}^K \binom{K}{\ell} (-1)^\ell a^\ell  =  (1-a)^K$. By applying the binomial expansion, $Q_0$ can be further simplified as follows:
\begin{eqnarray}
    Q_0 &\!\!\!=\!\!\!&  \frac{\varphi_0}{K(K-1)}  \sum\limits_{\ell = 0}^K \binom{K}{\ell} \left( \frac{(1+\epsilon_{_{\rm B}})(1+\epsilon_{_{\rm F}})}{P_{_{\rm F}}} \right)^{K-\ell} (-1)^\ell \left(\epsilon_{_{\rm B}}^{-1}  + 1  \right)^\ell   \int\nolimits_{\eta_{_{\rm B}}}^{\eta_{_{\rm B}}(1+\epsilon_{_{\rm F}})}    x^\ell dx  \nonumber \\
    &\!\!\!=\!\!\!&   \frac{\varphi_0 \epsilon_{_{\rm B}} (1+\epsilon_{_{\rm B}})^K}{P_{_{\rm F}}^{K+1}K(K-1)}   \sum\limits_{\ell = 0}^K \binom{K}{\ell} \frac{(-1)^\ell}{\ell+1}    \left( (1+\epsilon_{_{\rm F}})^{K + 1}  -  (1+\epsilon_{_{\rm F}})^{K-\ell}  \right).   \label{app:Q0}
\end{eqnarray} 

\subsection{High SNR approximation for $Q_k$ with $1 \le k \le K-2$}

Recalling the results in Appendix A, $Q_k$ can be rewritten as follows:
\begin{eqnarray}
    Q_k  &\!\!\!=\!\!\!& \varphi_k  \sum\limits_{n=0}^{K-k}   \binom{K\!-\!k}{n} (-1)^n \sum\limits_{\ell=0}^{k} \binom{k}{\ell} (-1)^{\ell}e^{\frac{\ell}{P_{_{\rm F}}}}  \mu_3 
    \int\nolimits_{\eta{_{\rm B}}}^{\eta{_{\rm B}}(1+\epsilon{_{\rm F}})}  e^{-\left( \frac{\ell}{P{_{\rm F}} \eta{_{\rm B}}} + \mu_4 +1 \right)x}  dx.  
\end{eqnarray}
By applying the approximation $e^{-x} \approx 1 - x$ for $x \to 0$ and using the definitions of $\mu_3$ and $\mu_4$, $Q_k$ can be approximated as follows:
\begin{eqnarray}
    Q_k  &\!\!\!=\!\!\!& \varphi_k  \sum\limits_{n=0}^{K-k}   \binom{K\!-\!k}{n} (-1)^n \sum\limits_{\ell=0}^{k} \binom{k}{\ell} (-1)^{\ell}e^{\frac{\ell}{P_{_{\rm F}}}}   e^{-\frac{ n \left(1+\epsilon_{_{\rm B}} \right) \left(1+ \epsilon_{_{\rm F}} \right) }{P_{_{\rm F}}}}
    \nonumber \\
    &\!\!\! \!\!\!& \times \int\nolimits_{\eta{_{\rm B}}}^{\eta{_{\rm B}}(1+\epsilon{_{\rm F}})}  e^{-\left( \frac{\ell}{P{_{\rm F}} \eta{_{\rm B}}} -  \frac{n}{P_{_{\rm F}} \eta_{_{\rm B}} }  - n  \right)x}  dx.  
\end{eqnarray}
By using the fact that $\sum\limits_{\ell=0}^{k} \binom{k}{\ell} (-1)^{\ell} a^\ell = (1-a)^k$, the above $Q_k$ can be expressed as follows:
\begin{eqnarray}
    Q_k  &\!\!\!=\!\!\!& \varphi_k \int\nolimits_{\eta_{_{\rm B}}}^{\eta_{_{\rm B}}(1+\epsilon_{_{\rm F}})}   \sum\limits_{n=0}^{K-k}   \binom{K\!-\!k}{n} (-1)^n \sum\limits_{\ell=0}^{k} \binom{k}{\ell} (-1)^{\ell}
    \nonumber \\
    &\!\!\! \!\!\!& \times  e^{-n\left(\frac{    (1+\epsilon_{_{\rm B}}  )  (1+ \epsilon_{_{\rm F}}  ) }{P_{_{\rm F}}} -(\epsilon_{_{\rm B}}^{-1} + 1)x \right)} e^{-\ell\left( \epsilon_{_{\rm B}}^{-1}x -  P_{_{\rm F}}^{-1}  \right)}  dx \nonumber \\
    &\!\!\! =  \!\!\!&  \varphi_k \int\nolimits_{\eta{_{\rm B}}}^{\eta{_{\rm B}}(1+\epsilon{_{\rm F}})} \left( 1 -  e^{- \left(\frac{    (1+\epsilon_{_{\rm B}}  )  (1+ \epsilon_{_{\rm F}}  ) }{P_{_{\rm F}}} -(\epsilon_{_{\rm B}}^{-1} + 1)x \right)}  \right)^{K-k}   \left( 1 -  e^{- \left( \epsilon_{_{\rm B}}^{-1}x -P_{_{\rm F}}^{-1}   \right)} \right)^{k}   dx  \nonumber \\
    &\!\!\! \mathop =\limits^{(a)}  \!\!\!&  \varphi_k \int\nolimits_{\eta{_{\rm B}}}^{\eta{_{\rm B}}(1+\epsilon{_{\rm F}})}    \left(\frac{    (1+\epsilon_{_{\rm B}}  )  (1+ \epsilon_{_{\rm F}}  ) }{P_{_{\rm F}}} -(\epsilon_{_{\rm B}}^{-1} + 1)x   \right)^{K-k}  \left(  \epsilon_{_{\rm B}}^{-1}x - P_{_{\rm F}}^{-1}   \right) ^{k}   dx ,  \label{apb:Q_k}
\end{eqnarray} 
where step (a) follows high SNR approximations. Applying the binomial expansions to \eqref{apb:Q_k}, the high SNR approximation for $Q_k$ can be obtained as follows:
\begin{eqnarray}
    Q_k &\!\!\! = \!\!\!&  \frac{\varphi_k \epsilon_{_{\rm B}}  (1+\epsilon_{_{\rm B}})^{K-k} (-1)^k }{ P_{_{\rm F}}^{K+1} } \sum\limits_{n=0}^{K-k} \binom{K-k}{n} (-1)^n (1 + \epsilon_{_{\rm F}} )^{K-k-n} \nonumber \\
    &\!\!\! \!\!\!&    \times  \sum\limits_{\ell=0}^{k} \binom{k}{\ell} (-1)^\ell     
    \frac{(1 + \epsilon_{_{\rm F}} )^{n+\ell+1} - 1}{n+\ell+1}.  \label{app:Qk}
\end{eqnarray}

\subsection{High SNR approximation for $Q_{K-1}$}

First, we rewrite $Q_{K-1}$ as follows:
\begin{eqnarray}
    Q_{K-1}  &\!\!\!=\!\!\!& \frac{\varphi_0}{K-1} \sum\limits_{\ell=0}^{K-1} \binom{K\!-\!1}{\ell}  (-1)^\ell e^{\frac{\ell}{P_{_{\rm F}}} }    \left( e^{\frac{1}{P_{_{\rm F}}}} \nu(\ell, \mu_5)  - e^{-\frac{\epsilon_{_{\rm B}} + \epsilon_{_{\rm F}} + \epsilon_{_{\rm B}}\epsilon_{_{\rm F}}  }{P_{_{\rm F}}} } \nu(\ell, \mu_6)   \right) \nonumber \\
    &\!\!\! =  \!\!\!& \frac{\varphi_0}{K-1} \sum\limits_{\ell=0}^{K-1} \binom{K\!-\!1}{\ell}  (-1)^\ell e^{\frac{\ell}{P_{_{\rm F}}} }  \int\nolimits_{\eta{_{\rm B}}}^{\eta{_{\rm B}}(1+\epsilon{_{\rm F}})}  \left(   e^{\frac{1}{P_{_{\rm F}}}}  e^{-\left( \frac{\ell}{P{_{\rm F}} \eta{_{\rm B}}} + \mu_5 +1 \right)x}  \right.  \nonumber \\
    &\!\!\!    \!\!\!& \left. - e^{-\frac{\epsilon_{_{\rm B}} + \epsilon_{_{\rm F}} + \epsilon_{_{\rm B}}\epsilon_{_{\rm F}}  }{P_{_{\rm F}}} }  e^{-\left( \frac{\ell}{P{_{\rm F}} \eta{_{\rm B}}} + \mu_6 +1 \right)x} \right)  dx. 
\end{eqnarray}
By using the fact that $\sum\limits_{\ell=0}^{k} \binom{k}{\ell} (-1)^{\ell}   = 0$, $Q_{K-1}$ can be expressed as follows:
\begin{eqnarray}
    Q_{K-1}  &\!\!\!=\!\!\!& \frac{\varphi_0}{K-1}   \int\nolimits_{\eta_{_{\rm B}}}^{\eta_{_{\rm B}}(1+\epsilon_{_{\rm F}})} \left( e^{\frac{1}{P_{_{\rm F}}} - (\epsilon_{_{\rm B}}^{-1} + 1)x}  -  e^{-\frac{\epsilon_{_{\rm B}} + \epsilon_{_{\rm F}} + \epsilon_{_{\rm B}}\epsilon_{_{\rm F}}  }{P_{_{\rm F}}} }  \right)  \left( 1 - e^{ - \ell \left(\epsilon_{_{\rm B}}^{-1} x - P_{_{\rm F}}^{-1}  \right)}     \right)^{K-1}    dx. 
\end{eqnarray}
By applying the approximations $e^{-x} = 1- x$ for $x \to 0$, $Q_{K-1}$ can be approximated as follows:
\begin{eqnarray}
    Q_{K-1}  &\!\!\!=\!\!\!& \frac{\varphi_0}{K-1}   \int\nolimits_{\eta_{_{\rm B}}}^{\eta_{_{\rm B}}(1+\epsilon_{_{\rm F}})}     \left( \frac{(1+\epsilon_{_{\rm B}})(1+\epsilon_{_{\rm F}})}{P_{_{\rm F}}}  - (\epsilon_{_{\rm B}}^{-1} + 1)x   \right)  \left(  \epsilon_{_{\rm B}}^{-1} x - P_{_{\rm F}}^{-1}   \right)^{K-1}    dx \nonumber \\
    &\!\!\!=\!\!\!& \frac{\varphi_0}{K-1}   \int\nolimits_{\eta_{_{\rm B}}}^{\eta_{_{\rm B}}(1+\epsilon_{_{\rm F}})}   \epsilon_{_{\rm B}}^{-(K-1)}   \left( \frac{(1+\epsilon_{_{\rm B}})(1+\epsilon_{_{\rm F}})}{P_{_{\rm F}}}  - (\epsilon_{_{\rm B}}^{-1} + 1)x   \right)    \left(x - \frac{\epsilon_{_{\rm B}}}{P_{_{\rm F}}}   \right)^{K-1} dx   \nonumber \\
    &\!\!\!=\!\!\!&  \frac{\varphi_0}{K-1}   \int\nolimits_{\eta_{_{\rm B}}}^{\eta_{_{\rm B}}(1+\epsilon_{_{\rm F}})}   \epsilon_{_{\rm B}}^{-(K-1)}     \frac{(1+\epsilon_{_{\rm B}})(1+\epsilon_{_{\rm F}})}{P_{_{\rm F}}}   \left(x - \frac{\epsilon_{_{\rm B}}}{P_{_{\rm F}}}   \right)^{K-1} dx   \nonumber \\
    &\!\!\! \!\!\!&  - \frac{\varphi_0}{K-1}   \int\nolimits_{\eta_{_{\rm B}}}^{\eta_{_{\rm B}}(1+\epsilon_{_{\rm F}})}   \epsilon_{_{\rm B}}^{-(K-1)}    (\epsilon_{_{\rm B}}^{-1} + 1)x      \left(x - \frac{\epsilon_{_{\rm B}}}{P_{_{\rm F}}}   \right)^{K-1} dx   \nonumber \\
    &\!\!\!=\!\!\!&   \frac{\varphi_0 \epsilon_{_{\rm B}} \epsilon_{_{\rm F}}^K (1 + \epsilon_{_{\rm B}})(1+ \epsilon_{_{\rm F}})  }{ P_{_{\rm F}}^{K+1}K(K-1)} -     
    \frac{ \varphi_0  \epsilon_{_{\rm F}}^K (\epsilon_{_{\rm B}}^{-1} + 1) (K(1+\epsilon_{_{\rm F}}) + 1) }{P_{_{\rm F}}^{K+1}K(K-1)(K+1)} .  
    \label{app:QKm1}
\end{eqnarray}

Following similar procedures as for deriving the approximation of $Q_k$ ($1 \le k \le K-1$), $Q_K$ and $Q_{K+1}$ can be approximated as 
\begin{eqnarray}
    Q_K = \frac{\epsilon_{_{\rm B}} \epsilon_{_{\rm F}}^{K+1} }{P_{_{\rm F}}^{K+1}(K+1) }  + \frac{\epsilon_{_{\rm F}}^K}{P_{_{\rm F}}^K}  - \frac{\epsilon_{_{\rm B}} \epsilon_{_{\rm F}}^K(1+\epsilon_{_{\rm F}})}{P_{_{\rm F}}^{K+1}}  \label{app:QK}
\end{eqnarray}
and 
\begin{eqnarray}
    Q_{K+1} =  \frac{\epsilon_{_{\rm F}}^K \left( (1 + \epsilon_{_{\rm B}})^{K+1} -1 \right) }{P_{_{\rm F}}^{K+1} (K+1)}  - \frac{\epsilon_{_{\rm F}}^K \left( (\epsilon_{_{\rm B}}(K+1) -1 )(1+\epsilon_{_{\rm B}})^{K+1} +1  \right) }{P_{_{\rm F}}^{K+2} (K+2) (K+1)}   ,   \label{app:QKp1}
\end{eqnarray}
respectively. 

By combing \eqref{app:Q0}, \eqref{app:Qk}, \eqref{app:QKm1}, \eqref{app:QK}, and \eqref{app:QKp1}, the high SNR approximation for $P_{\rm out}$ is obtained. 

\section*{Appendix C: A proof of Corollary 3}
\renewcommand{\theequation}{C.\arabic{equation}}
\setcounter{equation}{0}
\setcounter{subsection}{0}

Recalling the expression in \eqref{eq:Pout_single}, the outage probability experienced by the single GF user can be rewritten as follows:
\begin{eqnarray} 
    P_{\rm out}  &\!\!\!=\!\!\!& \mathop{\mathcal{E}}\limits_{ \eta_{_{\rm B}} < |h_{_{\rm B}}|^2  < \eta_{_{\rm B}} (1+\epsilon_{_{\rm F}}) }  \left\{   \Pr \left( |h_1|^2 < \frac{(1+\epsilon_{_{\rm F}})(1+\epsilon_{_{\rm B}})-(1 + P_{_{\rm B}}|h_{_{\rm B}}|^2 ) }{P_{_{\rm F}}}  \right)  \right\}  \nonumber \\
    &\!\!\! \!\!\!& + \mathop{\mathcal{E}}\limits_{ |h_{_{\rm B}}|^2 > \eta_{_{\rm B}}(1 + \epsilon_{_{\rm F}})  }  \left\{   \Pr \left( |h_1|^2 < \frac{\epsilon_{_{\rm F}}}{P_{_{\rm F}}}  \right)  \right\}   
    + \mathop{\mathcal{E}}\limits_{ |h_{_{\rm B}}|^2 < \eta_{_{\rm B}}  }  \left\{   \Pr \left( |h_1|^2 < \frac{\epsilon_{_{\rm F}}(1 + P_{_{\rm B}} |h_{_{\rm B}}|^2)}{P_{_{\rm F}}}  \right)  \right\} \!.~~~~~~
         \label{ap:Pout_single}
\end{eqnarray}  
By using the approximation $1 - e^{-x} = x $ for $x \to 0$, the high SNR approximation for the outage probability can be calculated as follows:
\begin{eqnarray} 
    P_{\rm out}  &\!\!\!\approx \!\!\!&  \int\nolimits_{\eta_{_{\rm B}}}^{\eta_{_{\rm B}} (1+\epsilon_{_{\rm F}})}  \frac{ e^{-x}(\epsilon_{_{\rm B}} + \epsilon_{_{\rm F}} + \epsilon_{_{\rm B}} \epsilon_{_{\rm F}}  - P_{_{\rm B}} x) }{P_{_{\rm F}}}  dx
    + \int\nolimits_{\eta_{_{\rm B}}(1 + \epsilon_{_{\rm F}})}^{\infty} \frac{\epsilon_{_{\rm F}} e^{-x}}{P_{_{\rm F}}}  dx \nonumber \\
    &\!\!\! \!\!\!&  + \int\nolimits_0^{\eta_{_{\rm B}}} \frac{\epsilon_{_{\rm F}} e^{-x} (1+ P_{_{\rm B}} |h_{_{\rm B}}|^2)}{P_{_{\rm F}}}  dx  \nonumber \\
    &\!\!\! \mathop{\approx}\limits^{(a)} \!\!\!& \frac{\epsilon_{_{\rm B}}\epsilon_{_{\rm F}}(\epsilon_{_{\rm F}}-\epsilon_{_{\rm B}}) }{P_{_{\rm F}}^2}  + \frac{\epsilon_{_{\rm F}}}{P_{_{\rm F}}}  -  \frac{\epsilon_{_{\rm B}}\epsilon_{_{\rm F}}(1+\epsilon_{_{\rm F}})}{P_{_{\rm F}}^2}  + \frac{\epsilon_{_{\rm B}}\epsilon_{_{\rm F}}(1+\epsilon_{_{\rm B}})}{P_{_{\rm F}}^2}  \nonumber \\
    &\!\!\! =  \!\!\!&  \frac{\epsilon_{_{\rm F}}}{P_{_{\rm F}}} ,         
\end{eqnarray}  
where step (a) is obtained by using the approximation $1 - e^{-x} = x $ for $x \to 0$ again. 

\section*{Appendix D: A proof of Corollary 4}
\renewcommand{\theequation}{D.\arabic{equation}}
\setcounter{equation}{0}
\setcounter{subsection}{0}

By using \eqref{eq:Pout_simplified0}, an upper bound of the outage probability achieved by the RSMA-SGF scheme can be obtained as follows:
\begin{eqnarray}
    P_{\rm out}  &\!\!\!=\!\!\!&  \Pr \left( E_0,  R_{_{{\rm II},K}} < \hat R_{_{\rm F}} \right)  +  \sum\limits_{k=1}^{K-1} \Pr \left( E_k,  R_{_{{\rm II},K}}  < \hat R_{_{\rm F}} \right)    +  \Pr \left( E_K,  R_{_{{\rm I},K}} < \hat R_{_{\rm F}} \right) \nonumber \\
    &\!\!\! \le \!\!\!&  \sum\limits_{k=0}^{K-1}  \underbrace{ \Pr \left(  |h_K|^2 > \frac{\tau}{P_{_{\rm F}}}, R_{_{{\rm II},K}}  < \hat R_{_{\rm F}}  \right) }_{Q_F}  + \Pr \left( R_{_{{\rm I},K}} < \hat R_{_{\rm F}} \right) \nonumber \\
    &\!\!\!=\!\!\!& K Q_F   + \Pr \left(   |h_K|^2 < \frac{\epsilon_{_{\rm F}} }{ P_{_{\rm F}} }   \right).   \label{eq:outage_floor}
\end{eqnarray}
By using the definitions of $\tau$ and $R_{_{{\rm II},K}}$, $Q_F$ can be rewritten as follows:
\begin{eqnarray}
    Q_F = \Pr \left(  |h_K|^2 > \frac{P_{_{\rm B}}\epsilon_{_{\rm B}}^{-1}|h_{_{\rm B}}|^2 - 1 }{P_{_{\rm F}}}, |h_K|^2 < \frac{(1+\epsilon_{_{\rm F}})(1+\epsilon_{_{\rm B}})-(1 + P_{_{\rm B}}|h_{_{\rm B}}|^2 ) }{P_{_{\rm F}}}   \right)  .
\end{eqnarray}
Obviously, the lower bound on $|h_K|^2$, i.e., $\frac{P_{_{\rm B}}\epsilon_{_{\rm B}}^{-1}|h_{_{\rm B}}|^2 - 1 }{P_{_{\rm F}}}$, needs to be smaller than the corresponding upper bound, i.e., $\frac{(1+\epsilon_{_{\rm F}})(1+\epsilon_{_{\rm B}})-(1 + P_{_{\rm B}}|h_{_{\rm B}}|^2 ) }{P_{_{\rm F}}}$, which introduces a hidden constraint 
\begin{eqnarray}
|h_{_{\rm B}}|^2 < \eta_{_{\rm B}} (1 + \epsilon_{_{\rm F}}). \label{eq:eta_B_constraint} 
\end{eqnarray}
Since the expression in \eqref{eq:eta_B_constraint} does not require any additional constraint on  $\epsilon_{_{\rm B}}$ and  $\epsilon_{_{\rm F}}$, $Q_F$ can be evaluated as follows:
\begin{eqnarray}
    Q_F &\!\!\!=\!\!\!&   \Pr \left(  |h_K|^2 > \frac{P_{_{\rm B}}\epsilon_{_{\rm B}}^{-1}|h_{_{\rm B}}|^2 - 1 }{P_{_{\rm F}}}, |h_K|^2 < \frac{(1+\epsilon_{_{\rm F}})(1+\epsilon_{_{\rm B}})-(1 + P_{_{\rm B}}|h_{_{\rm B}}|^2 ) }{P_{_{\rm F}}} , \right. \nonumber \\
    &\!\!\! \!\!\!& \left.  ~~~~~~   |h_{_{\rm B}}|^2 < \eta_{_{\rm B}} (1 + \epsilon_{_{\rm F}})   \Big) \right.  \nonumber \\
    &\!\!\! \le \!\!\!&   \Pr \left(  |h_{_{\rm B}}|^2 < \eta_{_{\rm B}} (1 + \epsilon_{_{\rm F}})  \right)   = 1 - e^{-  \frac{\epsilon_{_{\rm B}} (1 + \epsilon_{_{\rm F}})}{P_{_{\rm F}}}} ~  \mathop{\to}\limits^{{P_{_{\rm B}} \to \infty}} ~  0,
\end{eqnarray}
where the last inequality indicates that the constraint $|h_{_{\rm B}}|^2 < \eta_{_{\rm B}} (1 + \epsilon_{_{\rm F}}) $ effectively removes the outage floor. On the other hand, the remaining term $\Pr \left(   |h_K|^2 < \frac{\epsilon_{_{\rm F}} }{ P_{_{\rm F}} }   \right)$ in \eqref{eq:outage_floor} also approaches zero in the high SNR region. Therefore, without requiring any additional constraint on  the target rate values (equivalently $\epsilon_{_{\rm B}}$ and  $\epsilon_{_{\rm F}}$), $P_{\rm out}$ does not experience an outage floor in the RSMA-SGF scheme.  

\begin{balance}
\bibliography{IEEEabrv,IEEE_bib}
\end{balance}

\end{document}